%% file: ex_article.tex
\newif\ifarxiv
\arxivtrue

\makeatletter
\@ifundefined{AddToHook}{%
  \RequirePackage{etoolbox}%
  \long\def\AddToHook#1#2{\siamcompat@addtohook#1\@nil{#2}}%
  \long\def\siamcompat@addtohook env/#1/begin\@nil#2{%
    \AtBeginEnvironment{#1}{#2}%
  }%
}{}
\makeatother

\ifarxiv\else
  \PassOptionsToClass{review}{siamonline250211}
\fi

\documentclass[onefignum,onetabnum, nohypdvips]{siamonline250211}

\input{ex_shared}
\ifarxiv
  \title{Optimal Gating for Dead-Time–Constrained Single-Photon Detectors}

  \colorlet{header1}{black}
  \hypersetup{hidelinks}
  \renewcommand{\footercopyright}{}
  \renewcommand{\fj}[1]{}
  
  \makeatletter
  \renewcommand{\@maketitle}{%
    \newpage
    \null
    \vskip 1em
    \begin{center}
      {\LARGE\normalfont\bfseries \@title\par}
      \vskip 1em
      {\large\normalfont
        \def\and{\unskip\quad and\quad\ignorespaces}%
        \@author\par}
    \end{center}
    \vskip 1em
  }
  \renewcommand{\ps@plain}{%
    \let\@mkboth\@gobbletwo
    \let\@oddhead\@empty
    \let\@evenhead\@empty
    \def\@oddfoot{\normalfont\small\hfil\thepage\hfil}%
    \let\@evenfoot\@oddfoot
  }
  \makeatother
\fi
\usepackage{amssymb}
\usepackage{placeins}
\usepackage{needspace}

\ifpdf
\hypersetup{
  pdftitle={Optimal Gating for Dead-Time--Constrained Single-Photon Detectors},
  pdfauthor={Frederic J. N. Jorgensen and Steven G. Johnson}
}
\fi

\begin{document}
\raggedbottom % Keep natural spacing when an algorithm moves to the next page.

\maketitle

% REQUIRED
\begin{abstract}
We develop a Fisher-information framework for optimizing single-photon detector
gating in periodic experiments, in contrast to previous more empirical or heuristic gating choices. After each photon detection, such a detector is insensitive
for a \emph{dead time}, so that an early background detection can obscure a later signal. 
\emph{Gating} preserves availability for informative photons by selectively deactivating
the detector. In our high-pile-up simulation of distance-attenuated lidar, our practical data-independent gating design
reduces the worst-case asymptotic range variance by more than half relative to the 
free-running acquisition (which does not use any gating). More generally, we characterize achievable
detector-active frequencies as a convex polytope and construct an adaptive rule
that asymptotically attains an optimal ``oracle'' design without prior knowledge of the
physical parameters. We also derive practical criteria that quantify the potential gains of gating and/or photon attenuation over unattenuated free-running acquisition. 
\end{abstract}

% REQUIRED

% REQUIRED

\section{Introduction}
\begin{figure}[t]
  \centering
  \includegraphics[width=\linewidth]{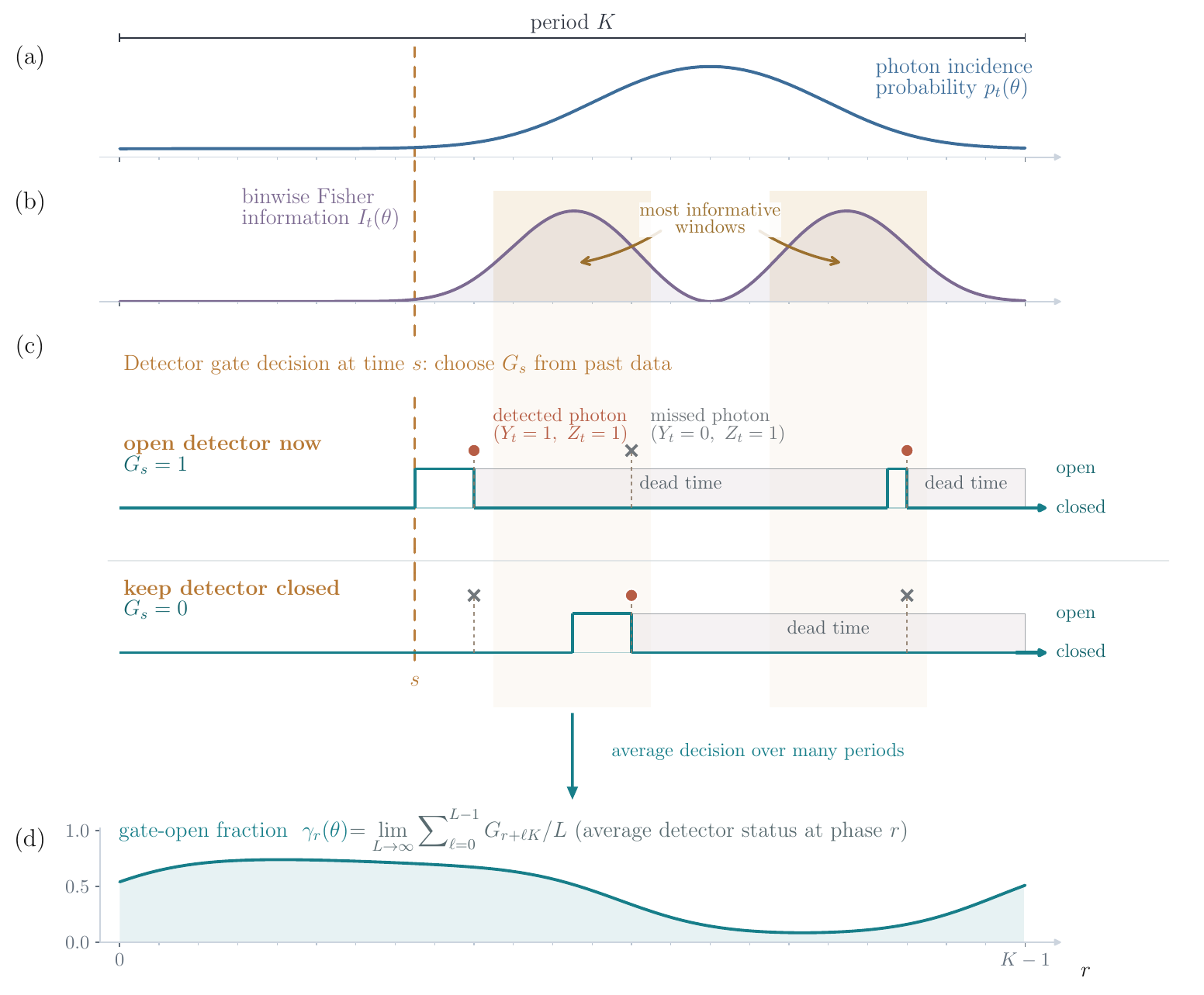}
  \caption{Conceptual overview of optimal gating design for a periodically illuminated
  single-photon experiment following notation introduced in \Cref{section:statistical_framework}. At a decision time $s$, the gating rule trades opening
  immediately against preserving detector availability for later (potentially more informative) times. Averaging
  such decisions over many periods yields the phasewise gating frequencies
  $\gamma_r(\theta)$ that enter our design objective.}
  \label{fig:conceptual_gating_design}
\end{figure}

Single-photon detectors are a key component of lidar (laser ranging) and many other
applications (\Cref{subsection:single_photon_imaging_examples}). Measurements of
photon arrival times relative to repeated excitation are used to estimate physical
parameters, such as the return-pulse delay in lidar
\cite{becker2005advanced,buller2009single,oconnor1984time}. 
After each detection, however, the detector becomes temporarily insensitive. 
This \emph{dead time} suppresses subsequent detections and distorts the
recorded arrival-time histogram, an effect known as \emph{pile-up}
\cite{muller1973dead,isbaner2016dead,rapp2019dead,li2017influence}.
In applications such as lidar (\Cref{subsection:single_photon_imaging_examples}), for example, an early background noise-photon detection
can therefore leave the detector unavailable when the return pulse arrives. Such dead-time distortion raises the question of which detector acquisition strategy is best for a given application and environment, with the goal of improving accuracy or, equivalently, achieving the same accuracy with less data.
One possible choice is \emph{free-running} mode, in which the detector is active as often as possible and re-activated
as soon as it has recovered from dead time \cite{rapp2019dead,isbaner2016dead}. 
Alternatively, in order to mitigate missing informative photons, \emph{gating} deliberately deactivates the detector during uninformative periods~\cite{koerner2024photon} to preserve its availability for later, more informative measurements
\cite{buttafava2014time,buttafava2015non}
(\Cref{fig:conceptual_gating_design}; \Cref{subsection:related_work}).  Another approach is to deliberately attenuate the incoming light, trading off reduced pile-up against the loss of incident photons~\cite{gupta2019photon}. Although such choices have previously been largely empirical, there is now an opportunity for a more principled approach: our recent work has identified fundamental accuracy bounds, optimal inference algorithms, and information metrics for such \textbf{d}ead-time \textbf{e}vent \textbf{d}etection (DED) problems~\cite{jorgensen2026fundamental}, providing a foundation on which to rigorously evaluate and design the detector strategy.

In this paper, we show how our recent information bounds~\cite{jorgensen2026fundamental} (reviewed in \Cref{section:statistical_framework}) can be applied to compare and optimize strategies for single-photon detector operation.
We bound the best possible improvement from gating (\Cref{section:gating_opt_problem}), yielding practical criteria
for when free-running is sufficient (\Cref{section:fr_opt}): we show that, among pre-planned gating sequences, free-running is optimal for sufficiently low pile-up and when the ensemble of signals is time-invariant (e.g., neglecting distance attenuation in lidar). To address more general scenarios, we then develop an optimization framework for gating design that can be applied to choose the parameters of any available implementation scheme (\Cref{section:opt_acq_design}).
In particular, we apply our framework to distance-attenuated lidar (\Cref{subsec:lidar}) over ranges of $2$--$5\,\mathrm m$,
optimizing over a hardware-compatible acquisition family. In our high-pile-up
numerical example, this pre-planned rule roughly halves the worst-case asymptotic
range-variance coefficient relative to free-running,  which allows for a higher update rate at the same estimation accuracy.  Finally, we show that little or no further improvement in accuracy arises from attenuating incoming light in addition to gating (\Cref{app:attenuation}).

In more technical terms, the underlying dead-time event-detection model and Fisher-information theory
provide our asymptotically optimal estimation-error objective
(\Cref{section:statistical_framework}) in terms of a vector $\gamma$ of gating frequencies (the detector-active fractions at each time, or ``phase'', within the period of the repeated experiments).  We characterize the achievable $\gamma$ (that is, consistent with the dead time) as a convex polytope and show that an adaptive rule can asymptotically attain the optimum of the resulting convex optimization problem without knowing the physical parameters (\Cref{section:gating_opt_problem}).
Comparison with this optimal solution gives a sharp bound on the regret of free
running and an exact optimality condition among all causal rules. 
Under a cyclic-shift symmetry of the underlying physical models (such as an unattenuated lidar pulse), free-running also minimizes the average
and worst-case delay losses among stationary ergodic schedules planned
independently of the photon data whenever the pile-up $\rho$, the largest expected
photon count in a dead-time window, is at most $\log 2$ uniformly over delay
(\Cref{section:fr_opt}). Minimizing our objective, the trace of the inverse Fisher information, is the A-optimality criterion of optimal experimental design~\cite[Section~7.5.2]{boyd2004convex}.

\subsection{Examples of single-photon imaging with dead time}
\label{subsection:single_photon_imaging_examples}

Time-correlated single-photon counting (TCSPC), often implemented with
single-photon avalanche diodes (SPADs) which are subject to dead time, uses arrival times relative to
repeated excitation to infer physical parameters of interest
\cite{becker2005advanced,buller2009single,oconnor1984time}. In lidar, our main example,
the return pulse's delay encodes the distance to an illuminated scene point
\cite{dong2017lidar,kirmani2014first}. In fluorescence lifetime imaging
(FLIM), the decay after excitation provides contrast between molecular
species \cite{becker2004fluorescence,becker2012fluorescence}.
Other related applications include non-line-of-sight imaging
\cite{rapp2020seeing,faccio2020non}, optical quantum information
processing \cite{hadfield2009single}, and X-ray astronomy
\cite{weisskopf2002overview,poppenhaeger2013transit}.
Dead-time effects on single-photon estimation have also been studied in non-line-of-sight ultraviolet
communications \cite{drost2015dead}, X-ray timing
\cite{tuo2024revisiting}, and radiation measurements
\cite{knoll2010radiation,gesualdi2022estimation}. 
Our framework applies to all periodic experiments satisfying the observation
model presented in \Cref{section:statistical_framework}.

\subsection{Related work}
\label{subsection:related_work}
At low photon flux, dead-time effects are small and, therefore, often neglected
at the inference stage \cite{oconnor1984time,kirmani2014first,altmann2016lidar}. When photon flux is high, computational correction methods can account for the resulting pile-up in single-photon
records \cite{coates1968correction,isbaner2016dead,gupta2019photon,
pediredla2018signal,rapp2019dead,kitichotkul2023role,kitichotkul2025free,jorgensen2026fundamental}.

Moreover, when signal arrival times are approximately known, timed gating
can reject strong early noise returns and extend the measurement's dynamic range
\cite{buttafava2014time,buttafava2015non,tosi2011fast,pifferi2008time}.  Another such acquisition scheme is the free-running scheme \cite{isbaner2016dead,rapp2019dead} in which the detector is in a sensitive state whenever possible. Representative acquisition families also include windows placed near expected
signal arrivals \cite{buttafava2014time,buttafava2015non} and gates shifted
across repeated periods \cite{wang1991two,tosi2011fast,reilly2014high,
gupta2019asynchronous}. For the case of single-photon lidar, Gupta et al.~\cite{gupta2019asynchronous} motivate uniform gate shifting by
minimizing an approximate upper bound on the average probability of
selecting the wrong depth bin.
The adaptive method of
\citeasnoun{po2022adaptive} uses Thompson sampling to adapt a single activation phase within a first-photon
acquisition protocol, which waits after dead time for a new acquisition
cycle.

Our companion paper \cite{jorgensen2026fundamental} establishes asymptotically efficient inference under general acquisition/gating schemes. Its performance bounds, attained asymptotically by practical estimators, motivate comparing acquisition rules by the estimation error remaining after such efficient estimation. Our aim in this paper is to characterize the feasible acquisition space and compare gating rules under a common asymptotic estimation-error criterion. This separates the fundamental limits imposed by detector dead time from those imposed by
a chosen class of acquisition schemes. 

\subsection{Notation}
We write $\mathbb N=\{0,1,2,\dots\}$, $\R$ for the real numbers, and $\mathbb Z_K=\{0,\ldots,K-1\}$ with arithmetic modulo $K$. We write $\mathbf 1$ and $\mathbf 0$ for the all-ones and all-zeros
vectors, and inequalities between vectors are defined componentwise. For a symmetric matrix
$X$, $X\succ 0$ and $X\succeq 0$ denote positive definiteness and semidefiniteness, and
$\succeq$ more generally denotes the corresponding Loewner order. We write $\mathbb S_+^d$ for the real symmetric positive-semidefinite $d\times d$ matrices, $I_d$ for the identity matrix, and $\odot$ for componentwise multiplication of tensors. We write $\Tr$ for the trace, $\Diag(\cdot)$ for a
diagonal matrix, and $\|\cdot\|_1$ for both the $\ell_1$ vector norm and the induced $\ell_1$
operator norm. $\|\cdot\|_2$ and $\|\cdot\|_\infty$ denote the Euclidean and maximum vector norms. Convergence in $P_\theta$-probability is denoted
$\xrightarrow{P_\theta}$, and $o_P(1)$, $O_P(1)$ are the usual stochastic-order symbols. 
We write $\ind\{E\}$ for the indicator function of a condition or set $E$, equal to $1$ if $E$ holds and $0$ otherwise, and $\delta_{rj}$ for the Kronecker delta.

\section{Fundamental bounds of single-photon detection}
\label{section:statistical_framework}

We review the dead-time event detection (DED) model and the Fisher-information estimation
bounds of \citeasnoun{jorgensen2026fundamental}, which provide the objective used for gating design in the remainder of this paper.

\subsection{Modeling single-photon detection}
\label{subsection:modeling_single_photon_detection}

Following \citeasnoun{jorgensen2026fundamental}, we divide time into bins of width
$\Delta$, indexed by $t\in\mathbb N$, with an illumination period of $K$ bins, where $K\ge1$ is an integer. This
discretization matches finite-resolution histogram readout in TCSPC
systems~\cite{zhang201830,zhang2021240,
kitichotkul2025free,wang2026synchronous}.
In the semiclassical picture that we consider here, incident photon counts follow a Poisson model~\cite{goodman2015statistical,
snyder2012random}. In a given time bin with index $t$, a single-photon detector measures whether at least one
photon triggers an avalanche, giving the binary random variable
\begin{equation}
\label{eq:bernoulli_experiments}
Z_t\sim \Ber(p_t(\theta)),
\qquad
p_t(\theta)=1-\exp[-\lambda_t(\theta)]\,.
\end{equation}
The variables \(Z_t\) are independent across time. The expected count per bin
\(\lambda_t(\theta)\) accounts for attenuation, dark counts, and ambient background
and is \(K\)-periodic, \(\lambda_t(\theta)=\lambda_{t+K}(\theta)\). Here \(p_t(\theta)\)
is the probability of at least one arrival in bin \(t\), while the parameter
\(\theta\in\Theta\subseteq\R^d\) is the physical quantity one wishes to estimate. In single-photon
lidar, for example, the mean count is a shifted return pulse on a constant background,
\begin{equation}
\label{eq:lidar_model}
\lambda_r(\theta)=a f_r(\tau)+b,\qquad \theta=(a,\tau,b)\,,
\end{equation}
where \(a\) is the return strength, \(b\) the expected background count per bin, and the time-of-flight \(\tau\)
determines the range through \(R=c\tau/2\), with \(c\) the speed of light in the
medium~\cite{kirmani2014first, dong2017lidar, rapp2019dead}. A normalized binned wrapped Gaussian pulse is
\[
f_r(\tau)=\Pr\!\left\{
(\tau+\sigma X)\bmod t_r\in[r\Delta,(r+1)\Delta)
\right\},\qquad
X\sim\mathcal N(0,1),\quad t_r=K\Delta\,.
\]
Here $\sigma$ is the pulse width, $a$ is the expected signal
count per period, and $b$ is the expected background count per bin.

Let the known gating variable $G_t\in\{0,1\}$ indicate whether the detector is active. The \textit{recorded} detection is
\begin{equation}
\label{eq:observed_experiment}
Y_t=G_tZ_t\,.
\end{equation}
The main physical reason that not all incident events can be observed is dead time: following a
detection, the detector must be quenched and reset before another photon can be recorded.  We
write \(t_d\) for the physical dead time and \(D\in\mathbb N\) for the corresponding number of bins, taking \(t_d=D\Delta\) in this discretized model. In the
nonparalyzable model used here, this recovery interval has length \(D\)
bins~\cite{muller1973dead,buller2009single,isbaner2016dead,rapp2019dead,oconnor1984time}, so
\begin{equation}
\label{eq:dead_time}
Y_r=1 \implies G_s=0
\quad\text{for }s=r+1,\ldots,r+D\,.
\end{equation}
This creates the basic gating tradeoff.  Opening the detector at time \(t\) may produce an immediate detection, but that detection prevents observations in bins \(t+1,\ldots,t+D\).  Closing the detector at an uninformative time can therefore preserve availability for later bins that carry more information about \(\theta\).  The \emph{gating rule} is this causal design choice.  We formalize the resulting observation processes using the DED model of \citeasnoun{jorgensen2026fundamental}.
\begin{definition}[DED process]
\label{def:DED}
A stochastic process \(\{(G_t,Y_t)\}_{t\ge0}\), with \(G_t,Y_t\in\{0,1\}\), is a dead-time event detection (DED) process if the following conditions hold.
\begin{itemize}
    \item There exist independent Bernoulli incident events \(Z_t\sim\Ber(p_t(\theta))\) with \(p_{t+K}(\theta)=p_t(\theta)\) such that \(Y_t=G_tZ_t\).
    \item The dead-time constraint \eqref{eq:dead_time} holds.
    \item  $G_t$ is chosen before $Z_t$ from the past observations \((G_0,Y_0),\ldots,(G_{t-1},Y_{t-1})\), using parameter-independent randomization independent of the incident sequence.
    \item For every \(\theta\in\Theta\) and phase \(r=0,\ldots,K-1\), there is a
    \emph{gating frequency} \(\gamma_r(\theta)\in[0,1]\), meaning the limiting fraction of
    periods in which phase \(r\) is active, such that
\begin{equation}
\label{eq:phase_convergent}
\frac1L\sum_{\ell=0}^{L-1}G_{r+\ell K}
\xrightarrow{P_\theta}
\gamma_r(\theta),
\qquad L\to\infty\,.
\end{equation}
\end{itemize}
\end{definition}

We give two examples of gating schemes.  In the \emph{free-running} scheme, the detector is opened whenever dead time permits it:
\[
G_t^{\mathrm{fr}}=1-\sum_{u=1}^D Y_{t-u}\,,
\]
with the convention \(Y_s=0\) for \(s<0\).  In the \emph{synchronous} scheme, the detector is activated at the beginning of a period unless it is still in dead time, and then deactivated after the first detection in that period.  

The gating frequencies record how often each phase is sampled after both gating and dead time.  The existence of these phasewise gating frequencies is a mild assumption, satisfied by all schemes we consider in this paper on their stated parameter domains. We refer to \citeasnoun{jorgensen2026fundamental} for a detailed discussion of this assumption and a proof that the free-running and synchronous schemes satisfy it. For other schemes considered in this paper, the existence of gating frequencies is established in \Cref{lem:controller_balance,thm:adaptive_oracle,lem:exogenous_gating_balance}.

\subsection{Fundamental estimation bounds}
\label{subsection:fundamental_estimation_bounds}
For a given gating rule, the inference task is to estimate $\theta$ from $\{(G_t,Y_t)\}_{t=0}^{n-1}$. The Fisher information for a single sampled Bernoulli experiment at phase $r$ is 
\begin{equation}
\label{eq:Ir_def}
\mathcal I_r(\theta)
=
\frac{1-p_r(\theta)}{p_r(\theta)}
\nabla\lambda_r(\theta)\nabla\lambda_r(\theta)^\top\,.
\end{equation}
The Fisher information rate of an observation process with gating frequencies \(\gamma\in[0,1]^K\) is then
\begin{equation}
\label{eq:Irate_def}
\mathcal I(\theta;\gamma)
=
\frac1K\sum_{r=0}^{K-1}\gamma_r\mathcal I_r(\theta)\,,
\end{equation}
and for the true process $\theta = \theta_0$ we write
\begin{equation}
\label{eq:I0_def}
\mathcal I_0=\mathcal I(\theta_0;\gamma(\theta_0))\,.
\end{equation}
Importantly, in \eqref{eq:Irate_def}, the physics and inference task enter through \(\mathcal I_r(\theta)\), while the acquisition policy for \(G_t\) and dead-time effects enter only through the limiting gating frequencies \(\gamma_r\).

\begin{assumption}
\label{ass:lan}
Assume $\Theta$ is compact, $\theta_0\in\operatorname{int}\Theta$, $\mathcal I_0\succ0$.
Each $\lambda_r$ is positive on $\Theta$ and extends to a $C^3$
function on a neighborhood of $\Theta$. Further assume that if $\lambda_r(\theta)=\lambda_r(\theta_0)$ whenever
$\gamma_r(\theta_0)>0$, then $\theta=\theta_0$. 
\end{assumption}
Under \Cref{ass:lan}, the DED experiment is locally asymptotically
normal with information $\mathcal I_0$
\cite[Proposition~2]{jorgensen2026fundamental}. Consequently, for a regular
estimator $\hat\theta_n$ based on $n$ time bins, the asymptotic lower bound
can be expressed informally as
\cite[Theorem~1 and Corollary~1]{jorgensen2026fundamental}
\[
\mathbb E_{\theta_0}\!\left[\|\hat\theta_n-\theta_0\|_2^2\right]
\;\gtrsim\;\frac{1}{n}\Tr(\mathcal I_0^{-1})\,.
\]
The same quantity is locally asymptotic minimax over arbitrary estimators
\cite[Theorem~2]{jorgensen2026fundamental}. The maximum likelihood
estimator and suitable one-step estimators attain the corresponding
efficient Gaussian limit
\cite[Theorems~3 and~4]{jorgensen2026fundamental}.
These results motivate minimizing $\Tr(\mathcal I_0^{-1})$ over feasible gating frequencies.
Throughout, we assign loss $+\infty$ to singular information matrices.  Note that the $\Vert \theta \Vert_2$ norm and the Fisher information $\mathcal{I}$ depend implicitly on the choice of units of the parameters $\theta$; changing units (or other changes of variables) is equivalent to a differently weighted A-optimality.

\section{The optimal-gating problem}
\label{section:gating_opt_problem}

The preceding section motivates designing gating rules by minimizing
$\Tr(\mathcal I(\theta_0;\gamma)^{-1})$ over feasible gating frequencies $\gamma \in [0,1]^K$ which we characterize
here as a convex polytope, over which minimizing $\Tr(\mathcal I(\theta_0;\gamma)^{-1})$ is a convex optimization problem (\Cref{section:relate_gating_vectors}). This objective is convex in $\gamma$ because $\mathcal I(\theta_0;\gamma)$ is linear in $\gamma$ and $X\mapsto\Tr(X^{-1})$ is convex on the positive-definite cone~\cite[Exercise~3.18(a)]{boyd2004convex}. We also construct an ``oracle'' rule, whose objective value we prove is asymptotically attained by an adaptive algorithm that adjusts the gating in response to measurements. More practical versions of this design problem are treated in \Cref{section:opt_acq_design}.

\subsection{Achievable gating frequencies}
\label{section:relate_gating_vectors}
Without dead time, the ideal choice would be $\gamma=\mathbf1$: always observing
every phase provides the most information. Indeed, increasing any
$\gamma_r$ adds a positive-semidefinite contribution to
$\mathcal I(\theta_0;\gamma)$ and therefore cannot increase
$\Tr(\mathcal I(\theta_0;\gamma)^{-1})$.
Dead time generally rules out this choice. A detection at one phase
prevents observations during the next $D$ bins, so the frequencies
$\gamma_r$ cannot be chosen arbitrarily within $[0,1]$.
We must therefore determine which vectors $\gamma$ can actually be produced by
a causal gating rule. The following theorem gives linear constraints
that are both necessary and sufficient, identifying the achievable
frequencies as a convex polytope.
\begin{theorem}
\label{thm:gating_constraints}
Fix $\theta_0\in \Theta$ with $p_r(\theta_0)\in[0,1)$ for every phase, and let $D\ge0$ be the dead time in bins. Let $A(\theta_0)\in\mathbb R^{K\times K}$ be defined through
\[
(A(\theta_0)\gamma)_r = \gamma_r + \sum_{u=1}^{D} p_{r-u}(\theta_0)\gamma_{r-u},
\qquad r=0,\dots,K-1\,,
\]
with all indices taken modulo $K$. Then every DED process with gating frequencies $\gamma(\theta_0)\in \R_{\geq 0}^K$ satisfies
$$
A\gamma \leq \mathbf{1}\,.
$$
Conversely, for every $\gamma \in \R_{\geq 0}^K$ with $A\gamma \leq \mathbf{1}$ there exists a DED process whose gating frequencies converge to $\gamma$ under $P_{\theta_0}$.
\end{theorem}
The converse is constructive and realized, for example, by the $\alpha$-scheme of \Cref{alg:alpha_scheme}, which opens an available detector at phase $r$ with a fixed probability $\alpha_r$. To describe it, we use the dead-time counter $R_t\in\{0,\dots,D\}$, defined recursively by
\begin{equation}
    \label{eq:dead_time_counter}
    R_{t+1}=D\,Y_t+(R_t-1)\ind\{R_t>0\},\qquad R_0=0\,,
\end{equation}
and the availability indicator
\begin{equation}
\label{eq:availability_indicator}
C_t:=\ind\{R_t=0\}\,,
\end{equation}
which equals $1$ exactly when the detector can register a detection.
\begin{algorithm}[H]
\caption{$\alpha$-scheme}
\label{alg:alpha_scheme}
\begin{algorithmic}[1]
\Require Opening probabilities $\alpha\in[0,1]^K$
\For{$t=0,1,2,\ldots$}
  \State Draw $U_t\sim\mathrm{Unif}[0,1]$ independently 
  \State Set $G_t\gets C_t\ind\{U_t\le\alpha_{t\bmod K}\}$ 
  \State Observe $Y_t=G_tZ_t$
\EndFor
\end{algorithmic}
\end{algorithm}
To realize a gating frequency vector $\gamma$ satisfying the hypothesis $A\gamma\le\mathbf 1$ of the theorem, one can run the $\alpha$-scheme of \Cref{alg:alpha_scheme} with opening probabilities $\alpha_r=\gamma_r/s_r$, where $s_r:=1-\sum_{u=1}^{D}p_{r-u}\gamma_{r-u}$ (feasibility implies $s_r>0$). The proof of \Cref{thm:gating_constraints} in \Cref{app:proof_gating_constraints} shows that this choice indeed attains $\gamma$.  

The feasibility condition can be rewritten as $\gamma_r\le 1-\sum_{u=1}^{D}p_{r-u}\gamma_{r-u}$ and formalizes the pile-up tradeoff. The long-run detection frequency at an earlier phase $r-u$ is $p_{r-u}\gamma_{r-u}$, and each such detection blocks phase $r$ within its dead-time window, so sampling one phase more can reduce availability at later phases. The following corollary shows that the free-running scheme, which opens the detector whenever it is available, makes every such constraint tight.
\begin{corollary}
\label{cor:free-running}
The free-running scheme is a DED process with gating frequencies $\gamma^{\mathrm{fr}}$ satisfying $A\gamma^{\mathrm{fr}}=\mathbf{1}$.
\end{corollary}
The proof is given in \Cref{app:proof_free_running}.
Combining \Cref{thm:gating_constraints} with the error-rate objective of \Cref{subsection:fundamental_estimation_bounds}, optimal gating design becomes the convex optimization problem
\begin{equation}
\label{eq:gating_convex_problem}
\gamma_\star \in \arg\min_{\substack{\gamma\ge\mathbf 0\\ A(\theta_0)\gamma\le\mathbf 1}}
\Tr\!\big(\mathcal I(\theta_0;\gamma)^{-1}\big)\,.
\end{equation}
This is an A-optimal design problem~\cite{pronzato2008optimal,boyd2004convex,herzog2020first}.
We study when the free-running scheme (approximately) solves this optimization problem in \Cref{section:fr_opt}.

\subsection{Oracle and adaptive gating}
\label{section:adaptive_algorithm}
The convex optimization problem~\eqref{eq:gating_convex_problem} at the true parameter
$\theta_0$ defines the \emph{oracle} gating, the optimum available to an experimentalist who already knew $\theta_0$ in advance. In an actual experiment, one aims to estimate $\theta_0$ and does not know it in the first place, but an adaptive algorithm can asymptotically attain the oracle information objective by adjusting the gating scheme as data accumulates. In particular, we show that
\Cref{alg:adaptive_plugin}, which uses an improving pilot estimate to tune the
$\alpha$-scheme, yields gating frequencies attaining the optimum in \eqref{eq:gating_convex_problem} without knowing $\theta_0$ in advance (\Cref{thm:adaptive_oracle}).

Let $\tilde U\subseteq\operatorname{int}\Theta$ be an open subset of the parameter space.

\begin{algorithm}[!htbp]
\caption{Adaptive gating}
\label{alg:adaptive_plugin}
\begin{algorithmic}[1]
\Require Opening maps $\alpha^{(m)}:\tilde U\to[0,1]^K$, $m\ge1$
\State Set $L_{\mathrm{pil}}\gets K(\lceil D/K\rceil+1)$, $T_1\gets0$, and $R_0\gets0$
\For{$m=1,2,\ldots$}
  \State Set $E_m\gets T_m+L_{\mathrm{pil}}$
  \State Draw $J_m\sim\mathrm{Unif}\{0,\ldots,K-1\}$ independently
  \State In $[T_m,E_m)$, open only at $T_m+L_{\mathrm{pil}}-K+J_m$ and record its outcome $X_m$
  \State Compute the maximum likelihood estimator (MLE)
  \[
  \hat\theta_m\in\arg\max_{\eta\in\Theta}
  \sum_{j=1}^{m}\left[
  X_j\log p_{J_j}(\eta)
  +(1-X_j)\log(1-p_{J_j}(\eta))
  \right]
  \]
  \If{$\hat\theta_m\in\tilde U$}
    \State Set $\alpha_m\gets\alpha^{(m)}(\hat\theta_m)$
  \Else
    \State Set $\alpha_m\gets0$
  \EndIf
  \State Starting at $E_m$, run the $\alpha_m$-scheme for $m$ periods
  \State Set $T_{m+1}\gets E_m+mK$
\EndFor
\end{algorithmic}
\end{algorithm}

\begin{theorem}
\label{thm:adaptive_oracle}
Suppose \Cref{ass:lan} holds at every $\theta\in\tilde U$
when all phases are included ($\gamma=\mathbf1$).
Then there exists a choice of opening maps $\alpha^{(m)}$ in
\Cref{alg:adaptive_plugin} such that, for every $\theta\in\tilde U$,
the phasewise gating frequencies converge in $L^1(P_\theta)$ to an optimizer of
\eqref{eq:gating_convex_problem} evaluated at $\theta_0=\theta$.
\end{theorem}
The construction and proof are given in \Cref{app:proof_adaptive_oracle}.
\begin{remark}
\label{rem:adaptive_target}
Under the same assumptions, let $\gamma:\tilde U\to[0,1]^K$ be continuous
and satisfy $A(\theta)\gamma(\theta)\le\mathbf1$ throughout $\tilde U$.
The same argument gives phasewise convergence to $\gamma(\theta)$ in
$L^1(P_\theta)$ when \Cref{alg:adaptive_plugin} uses any consistent pilot and
\[
\alpha^{(m)}_r(\eta)=\alpha_r(\eta):=
\frac{\gamma_r(\eta)}{1-\sum_{u=1}^D p_{r-u}(\eta)\gamma_{r-u}(\eta)},
\qquad \eta\in\tilde U\,.
\]
\end{remark}

While \Cref{alg:adaptive_plugin} can attain the optimal gating frequencies, it
requires online parameter estimation, which is demanding for high-rate acquisition
hardware. For this reason, we mostly consider this a constructive
attainability result for the optimum of \eqref{eq:gating_convex_problem}, against which other potentially more practical schemes that we consider in \Cref{section:opt_acq_design} can be compared.

\section[When is free-running (near-)optimal?]{When is free-running (near-)optimal?\nopunct}
\label{section:fr_opt}
Free-running acquisition, which opens the detector whenever dead time allows,
is widely used in single-photon imaging and ranging~\cite{isbaner2016dead,rapp2019dead,kitichotkul2025free}. This section
quantifies its gap to the optimum of~\eqref{eq:gating_convex_problem} and identifies when that gap vanishes, giving theoretical criteria for when gating design (for example as presented in \Cref{section:opt_acq_design}) can be practically useful. 

An important quantity for this analysis is \textit{pile-up}, the largest expected number of incident photons in a dead-time window, which we quantify  as 
\begin{equation}\label{eq:dead_time_photon_count}
\rho:=\max_r\sum_{u=1}^{D}\lambda_{r-u},
\qquad \lambda_r=-\log(1-p_r)\,,
\end{equation}
written with cyclic indices and $p_r=p_r(\theta_0)$ as defined in~\eqref{eq:bernoulli_experiments}. In \Cref{section:regret_and_optimality}, we quantify the regret and optimality in terms of pile-up compared to all causal schemes (including adaptive ones). In \Cref{subsection:q_block_gating}, we do a similar pile-up based analysis specifically for the case of unattenuated lidar and restricted to a more practical class of pre-planned gating schedules that are independent of the photon process, which allows us to prove a much stronger result. 

\subsection{On the regret and optimality of free-running acquisition}
\label{section:regret_and_optimality}
The potential improvement of \textit{any} acquisition scheme over the free-running one is upper bounded by pile-up~$\rho$ as the following result shows. 
\begin{theorem}
\label{thm:regret_via_lambda_minus}
Let $p_r\in(0,1)$, let $A$ be the constraint matrix of \Cref{thm:gating_constraints}, and suppose
\begin{equation}\label{eq:positivity_condition}
\sum_{r=0}^{K-1}\mathcal I_r\succ0\,.
\end{equation}
Write $\mathcal I^{\mathrm{fr}}:=\mathcal I(\theta;\gamma^{\mathrm{fr}})\succ0$ and let $m:=\min_r\gamma_r^{\mathrm{fr}}$. Every feasible $\gamma$ with positive-definite information satisfies
\begin{equation}\label{eq:fr_regret_matrix}
\mathcal I(\theta;\gamma)^{-1}\succeq m(\mathcal I^{\mathrm{fr}})^{-1}\,.
\end{equation}
At a fixed physical parameter $\theta$, let $f(\gamma)=\Tr(\mathcal I(\theta;\gamma)^{-1})$ denote our optimization objective.
Consequently, letting $f_\star:=\inf_{\gamma\ge0,\,A\gamma\le\mathbf1}f(\gamma)$ denote the optimum value, we obtain:
\begin{equation}\label{eq:fr_regret_sharp}
\frac{f(\gamma^{\mathrm{fr}})-f_\star}
{f(\gamma^{\mathrm{fr}})}
\le1-m\le1-e^{-\rho}\,.
\end{equation}
These bounds are sharp over the abstract class of periodic detection probabilities and information matrices. 
\end{theorem}
The proof of this result can be found in \Cref{app:proof_regret}. This result is intuitive and says that at low pile-up, no gating strategy can recover much information compared to free-running because detections rarely suppress subsequent detections.
\begin{remark}
Historically, the TCSPC literature suggested keeping detected counts between $1$--$5\%$ per cycle
\cite{gupta2019photon,pediredla2018signal,picoquant_tcspc} as a way to counter histogram pile-up, although inference methods that account for dead time can now support accurate estimation at higher photon fluxes (yielding higher data rates)~\cite{rapp2019dead,jorgensen2026fundamental}. At low flux, when $D\le K$ and the incident flux is concentrated within one dead-time window, $\rho$ approximately equals the detections per cycle.
Therefore, since $1-e^{-\rho}\approx\rho$ at low pile-up, the same rule then limits the possible acquisition-design gain to about $1$--$5\%$ by \Cref{thm:regret_via_lambda_minus}.
\end{remark}

\begin{remark}
In addition to gating, optical attenuation can  reduce pile-up at the
cost of discarding photons \cite{gupta2019photon}. However, we show in \Cref{prop:attenuation_gating} that attenuation adds little beyond oracle-optimal
gating at low per-bin photon counts, and no improvement when $D\ge1$ and
$\rho\le\log2$.
\end{remark}
We also characterize when free-running is optimal among all causal schemes by the exact criterion~\eqref{eq:fr_exact_certificate}, and give the intuitive sufficient condition~\eqref{eq:fr_local_certificate}.
\begin{proposition}
\label{prop:general_fr_optimal}
Let $p_r\in(0,1)$, let $A$ be the constraint matrix of \Cref{thm:gating_constraints}, and suppose~\eqref{eq:positivity_condition} holds.
Then $A$ is nonsingular, $\gamma^{\mathrm{fr}}=A^{-1}\mathbf1>\mathbf0$, and $\mathcal I^{\mathrm{fr}}:=\mathcal I(\theta;\gamma^{\mathrm{fr}})\succ0$. Define
\begin{equation}\label{eq:fr_information_weights}
h_r:=-\partial_r f(\gamma^{\mathrm{fr}})
=\frac1K\Tr\!\bigl((\mathcal I^{\mathrm{fr}})^{-2}\mathcal I_r\bigr),
\qquad \nu:=A^{-\top}h\,.
\end{equation}
Free-running minimizes $f$ over $\{\gamma\ge0:A\gamma\le\mathbf1\}$ if and only if
\begin{equation}\label{eq:fr_exact_certificate}
\nu\ge\mathbf0\,.
\end{equation}
In particular, the local inequalities
\begin{equation}\label{eq:fr_local_certificate}
p_r\sum_{u=1}^{D}h_{r+u}\le h_r,
\qquad r\in\mathbb Z_K\,,
\end{equation}
suffice. 
\end{proposition}
The proof is in \Cref{app:proof_general_fr_optimal}. Condition~\eqref{eq:fr_local_certificate} gives an interpretable ``balanced information'' criterion by comparing the immediate information value of opening with the expected value of the next $D$ bins lost after a detection. In lidar, however, these weights concentrate near the return pulse, so this local sufficient condition can easily fail. For such cases, we give a much stronger condition in the next section by restricting ourselves to a smaller class of gating schemes.

\subsection{Optimality in shift-symmetric settings among pre-planned schedules}
\label{subsection:q_block_gating}
For the case of unattenuated lidar (changing the delay shifts the return pulse without changing its
strength), we now analyze the optimality of free-running within the class of pre-planned schedules that we introduce below.  We show that if the maximum pile-up satisfies $\rho_\star\le\log2$, free-running
minimizes both the uniform-delay average and worst-case asymptotic estimation
error among these schedules (\Cref{thm:delay_optimality}).
\subsubsection{Pre-planned schedules}
\label{subsubsection:preplanned_schedules}
A planned gate $H_t\in\{0,1\}$ is a (possibly random) binary variable that specifies whether the detector should open during time bin $t$
if it is available. The actual gate and detection are then
\begin{equation}
\label{eq:block_rand_gating}
G_t=C_tH_t,
\qquad C_t=\ind\{R_t=0\},\qquad Y_t=G_tZ_t \, ,
\end{equation}
where $R_t$ is the dead-time counter \eqref{eq:dead_time_counter} and $C_t$ is the availability indicator \eqref{eq:availability_indicator}.
A \emph{pre-planned schedule} is a sequence of such planned gates
\[
H^{(\ell)}=(H_{\ell K},\ldots,H_{\ell K+K-1})\in\{0,1\}^K,
\qquad \ell\in\mathbb Z\,,
\]
independent of the entire photon process and with a law $\mathsf Q$ that does not depend
on the unknown parameters $\theta$. We only consider stationary and ergodic sequences of
such period gate vectors. \emph{Stationarity} means that the joint distribution of
$(H^{(\ell)},\ldots,H^{(\ell+m-1)})$ does not depend on $\ell$, for every block length $m\ge1$. \emph{Ergodicity} means that for every $m\ge1$ and $h\in(\{0,1\}^K)^m$,
\[
\frac1L\sum_{\ell=0}^{L-1}
\ind\{(H^{(\ell)},\ldots,H^{(\ell+m-1)})=h\}
\xrightarrow[L\to\infty]{\mathsf Q\text{-a.s.}}
\mathsf Q\bigl((H^{(0)},\ldots,H^{(m-1)})=h\bigr)\,.
\]
Thus the actual gates use no past-observation-dependent information beyond the detector's
dead time. Examples include free-running, the $\alpha$-scheme of
\Cref{alg:alpha_scheme} with opening probabilities fixed independently of the parameters,
and the suffix-randomized schedules of \Cref{section:opt_acq_design}.
Free-running has the point-mass law
$\mathsf Q^{\mathrm{fr}}=\delta_{(\mathbf1)_{\ell\in\mathbb Z}}$,
concentrated on the all-open sequence, so that $H_t\equiv1$ and $G_t=C_t$.

For every pre-planned schedule, the gating frequencies
$\gamma(\mathsf Q;p)$ of \eqref{eq:phase_convergent} exist
(\Cref{lem:exogenous_gating_balance}).
Together with the construction above, this establishes that $(G_t,Y_t)$ is a DED process
(\Cref{def:DED}).

\subsubsection{Unattenuated lidar problem}
\label{subsubsection:unattenuated_lidar}
We consider the unattenuated model here for which we use \eqref{eq:lidar_model} along $\theta_\tau=(a,\tau,b)$, holding $a,b$ fixed while $0\le\tau<t_r$. Changing $\tau$ therefore shifts the return pulse without changing its strength $a$. This is an idealization for the regime where range variation is small compared to the overall range, since the return strength $a$ of a fixed reflector decreases as range increases. Long-range lidar applications in which such idealizations are reasonable include depth imaging~\cite{chan2018long,li2019single}, terrestrial scanning~\cite{abellan2014terrestrial}, and cloud-boundary measurements~\cite{venema2000cloud}. We treat the attenuated case in \Cref{section:opt_acq_design}.

For a delay prior with density $\pi$ on $[0,t_r)$, define the averaged
and worst-case trace objectives
\begin{equation}
\label{eq:q_ranging_objectives}
\begin{aligned}
\mathcal R_{\mathrm{avg}}(\mathsf Q;\pi)
&:=\int_0^{t_r}\Tr\!\bigl(\mathcal I_\tau(\mathsf Q)^{-1}\bigr)\pi(\tau)\,d\tau\,,\\
\mathcal R_{\mathrm{wc}}(\mathsf Q)
&:=\sup_{\tau\in[0,t_r)}\Tr\!\bigl(\mathcal I_\tau(\mathsf Q)^{-1}\bigr)
\end{aligned}
\end{equation}
where the information rate \eqref{eq:Irate_def} entering the estimation-error objective of \Cref{subsection:fundamental_estimation_bounds} becomes
\[
\mathcal I_\tau(\mathsf Q)
:=\frac1K\sum_r\gamma_r(\mathsf Q;p(\tau))J_r(\tau)
\]
and $p_r(\tau)=1-e^{-\lambda_r(\theta_\tau)}$, and $J_r(\tau)=\mathcal I_r(\theta_\tau)$.

\subsubsection{Time-translation symmetry and optimality of free-running}
In the unattenuated model, shifting the delay by one bin cyclically relabels both the detection probabilities and the information matrices. Let $\rho(\omega)$ denote the pile-up at delay $\omega$. To cover all offsets within a bin, define
\begin{equation}\label{eq:uniform_delay_pileup}
\rho_\star:=\sup_{\omega\in[0,\Delta)}\rho(\omega)\,.
\end{equation}
\begin{theorem}
\label{thm:delay_optimality}
Let $t_r=K\Delta$ and consider a  delay model with $p_r(\tau)\in[0,1)$, $J_r(\tau)\in\mathbb S_+^d$, and $\sum_{r=0}^{K-1}J_r(\tau)\succ0$ for every $\tau\in[0,t_r)$. Suppose that, for $\omega\in[0,\Delta)$ and $j,r\in\mathbb Z_K$,
\[
p_r(\omega+j\Delta)=p_{r-j}(\omega),\qquad
J_r(\omega+j\Delta)=J_{r-j}(\omega)\,,
\]
as in the unattenuated lidar model of \Cref{subsubsection:unattenuated_lidar}. If $\rho_\star\le\log2$, then free running is optimal
\begin{equation}\label{eq:free_running_delay_optimality}
\begin{aligned}
\mathcal R_{\mathrm{avg}}(\mathsf Q^{\mathrm{fr}};1/t_r)
&=\min_{\mathsf Q}\mathcal R_{\mathrm{avg}}(\mathsf Q;1/t_r)\,,\\
\mathcal R_{\mathrm{wc}}(\mathsf Q^{\mathrm{fr}})
&=\min_{\mathsf Q}\mathcal R_{\mathrm{wc}}(\mathsf Q)\,,
\end{aligned}
\end{equation}
where $\mathsf Q$ ranges over stationary, ergodic pre-planned schedule laws. For a bounded prior  density $\pi$ with essential infimum $\pi_{\min}$ and essential supremum $\pi_{\max}$, the more general condition $\rho_\star\le\log(1+\pi_{\min}/\pi_{\max})$ gives optimality
\begin{equation}\label{eq:free_running_prior_optimality}
\mathcal R_{\mathrm{avg}}(\mathsf Q^{\mathrm{fr}};\pi)
=\min_{\mathsf Q}\mathcal R_{\mathrm{avg}}(\mathsf Q;\pi)\,.
\end{equation}
\end{theorem}
The proof follows from a more general information comparison under cyclic
shifts in \Cref{app:block_randomized}.

\begin{remark}
The results of \Cref{thm:regret_via_lambda_minus,prop:general_fr_optimal} also hold for any rescaled loss
$\Tr(W\mathcal I^{-1})$ with nonzero $W\succeq0$, with $h_r=-\partial_r f(\gamma^{\mathrm{fr}})$ computed for the chosen loss. The choices $W=I_d$ and $W=(c/2)^2e_\tau e_\tau^\top$ recover the joint trace loss and range-variance rate, respectively, where $e_\tau$ is the delay coordinate vector.  \Cref{thm:delay_optimality} also applies when there is uncertainty in $(a,b)$, provided the pile-up condition holds pointwise for every $(a,b)$ under consideration. 
\end{remark}

\begin{remark}
The threshold $\rho_\star\le\log2$ is sharp as a two-bin example establishes in \Cref{app:cyclic_threshold_sharp}.
\end{remark}

\subsection{Experiment illustrating the results}
\label{sec:exp_fr_regret}
We compare free-running with oracle-optimal gating for the unattenuated lidar
model of \Cref{subsection:modeling_single_photon_detection}, using the full-trace
loss for $\theta=(a,\tau,b)$, with $\tau$ measured in nanoseconds and $b$ the
background photons per bin. We use a $100\,\mathrm{ns}$ period with $0.1\,\mathrm{ns}$
bins ($K=1000$), a normalized wrapped-Gaussian pulse of width $1\,\mathrm{ns}$ at delay
$30\,\mathrm{ns}$, and $50\,\mathrm{ns}$ dead time ($D=500$).
These timing parameters are representative of those commonly used in single-photon lidar
simulations~\cite{rapp2019dead,gupta2019asynchronous,kitichotkul2025free,
arvani2018direct,incoronato2021statistical}. 
Code and data for reproducing all the numerical experiments in this paper are
available at \url{https://github.com/FredericJorgensen/optimal-gating}.

\begin{figure}[!t]
  \centering
  \resizebox{0.97\linewidth}{!}{\input{figure1_labels.tex}}
  \caption{Oracle-normalized regret of free-running acquisition for
  dead-time-constrained single-photon lidar. Panel (a) reports
  \(\Tr((\mathcal I^{\mathrm{fr}})^{-1})/\Tr(\mathcal I_\star^{-1})-1\)
  as a percentage, where \(\mathcal I_\star=\mathcal I(\theta;\gamma_\star)\),
  over signal counts \(a\) and background counts \(B=Kb\) per period.
  Contours mark constant regret. The blue region corresponds to \(\rho_\star\le\log2\), where
  free-running minimizes the uniform-average and worst-case losses
  over stationary, ergodic pre-planned schedules
  (\Cref{thm:delay_optimality}).
  Panel (b) compares \(\gamma_{\star,r}\) and \(\gamma_r^{\mathrm{fr}}\)
  at the marked point. The blue pulse profile is rescaled independently of
  the gating-frequency axis. Panel (c) plots the computed regret against
  \(\rho\) on logarithmic axes, varying \(B\) at each labeled signal strength
  \(a\), with \(\rho\le1.5\). The dashed curve is the upper bound \(e^\rho-1\)
  from \Cref{thm:regret_via_lambda_minus}, also shown as a percentage.}
  \label{fig:fr_regret}
  
\end{figure}
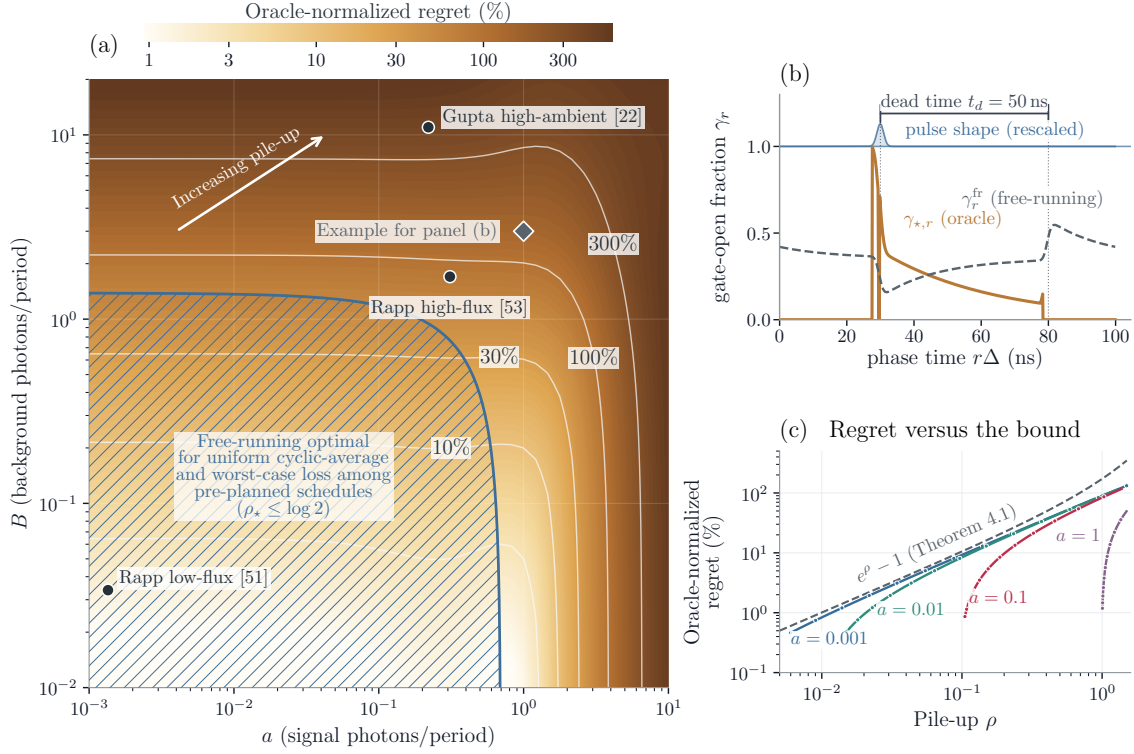
At each signal and background level in \Cref{fig:fr_regret}, we numerically
solve \eqref{eq:gating_convex_problem} and compare the best solution found with
free-running. The literature markers provide representative flux scales
\cite{rapp2017few,rapp2021high,gupta2019asynchronous}, evaluated under our common
timing model rather than reproducing the original experiments.

Free-running incurs little regret at low pile-up, while selective gating
can substantially improve performance at higher flux. Panel (b) illustrates
how suppressing detections before the return pulse preserves detector
availability around the pulse. Panel (c) shows that the computed regrets lie
below $e^\rho-1$, with a widening gap at larger pile-up. By contrast, the hatched region marks the parameter regime in which
\Cref{thm:delay_optimality} guarantees free-running optimality among
pre-planned schedules under delay uncertainty. We next study
distance-dependent attenuation, which breaks this shift symmetry.

\section{Designing practical gating schemes with application to single-photon lidar}
\label{section:opt_acq_design}
Adaptive acquisition can attain the oracle information objective (\Cref{section:adaptive_algorithm}),
but the necessary online parameter estimation can be complex to implement in hardware.
We therefore give a general method for optimizing the estimation criterion within
a prescribed class of schemes that, for example, much simpler, measurement-independent gating hardware can support. 
We then illustrate this method with the class of suffix-randomized gates for
distance-attenuated lidar, where attenuation breaks the shift symmetry of
\Cref{thm:delay_optimality}. In particular, we show over 50\% improvement over free-running in the high pile-up regime. 

\subsection{A practical design method}
\label{sec:practical_design_recipe}
\begin{figure}[htbp]
  \centering
  \includegraphics[width=\linewidth]{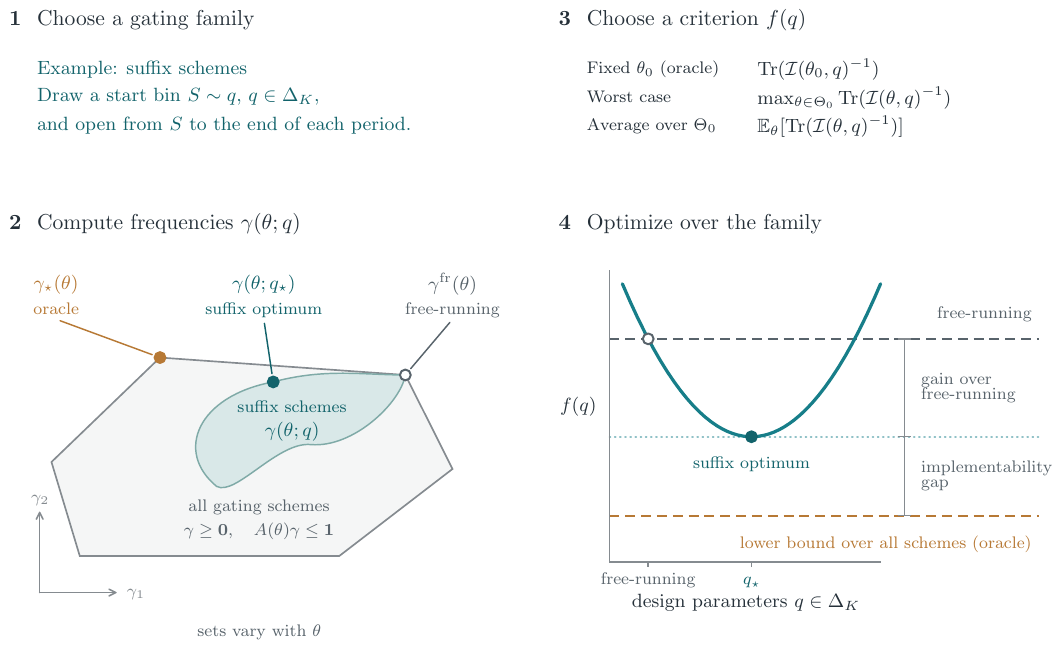}
  \caption{Practical gating design for the suffix family.
Panels 1--2 show the choice of a start law $q$ and computation of its phase frequencies
$\gamma(\theta;q)$ using \Cref{prop:suffix_gamma}. These frequencies
form a subset of $\{\gamma\ge0:A(\theta)\gamma\le\mathbf1\}$.
$q=\delta_0$ gives free running. Steps 3--4 choose an estimation criterion and optimize it over $q$. The pointwise oracle provides
a lower benchmark for these restricted designs.}

  \label{fig:design_recipe}
\end{figure}

Our proposed design method proceeds in four steps, also illustrated in \Cref{fig:design_recipe}.
\begin{enumerate}
\item Choose a class of rules $\{G_t\}$ satisfying \Cref{def:DED} parametrized by some design vector, so that the resulting  rules are feasible to implement on the available hardware.
\item Derive the induced gating frequencies $\gamma(\theta;\cdot)$ as a function of the design parameters and the physical parameter $\theta$.
\item Choose a design criterion built from the estimation rate $\Tr(\mathcal I(\theta;\gamma)^{-1})$, for example its average or its worst case over a range of plausible $\theta$.
\item Optimize the chosen criterion over the design parameters of the selected family.
\end{enumerate}
We carry out these steps for the example of suffix-randomized gates and attenuated lidar in \Cref{subsec:lidar}.

\subsection{Application to attenuated lidar}
\label{subsec:lidar}
We apply the suffix design method to \eqref{eq:lidar_model}, with
inverse-square decay in return strength and nominal operating points
\begin{equation}
\label{eq:attenuated_lidar_design_model}
\theta_R=
\left(S_\star(R_\star/R)^2,\;2R/c,\;B/K\right),
\qquad R\in[2,5]\,\mathrm m\,.
\end{equation}
For targets larger than the illuminated laser footprint, a standard return model
separates reflectivity from inverse-square spreading and exponential atmospheric loss \cite{jutzi2009normalization}. Here $S_\star$ absorbs fixed
reflectivity and system factors, and we neglect atmospheric loss over
$2$--$5\,\mathrm m$ to isolate geometric decay. Other target geometries or
propagation conditions may require a different range law, which the same
design method can accommodate. The range-dependent return strength breaks
the cyclic shift symmetry of \Cref{thm:delay_optimality}, allowing gains
over free-running.

\subsubsection{Step 1. A suffix-randomized rule}
We use \emph{block-randomized} pre-planned schedules
(\Cref{subsubsection:preplanned_schedules}) where the period vectors
$H^{(\ell)}$ are iid with a common law $Q$ on $\{0,1\}^K$.
These schedules require no online parameter estimation.
% , but their implementationmust respect the programmable gate's timing and waveform limits~\cite{scott2022multitime,mpd_ingaas_gated,mpd_pdmir}.
We restrict this family to suffix plans as follows. At the beginning of period $\ell$, draw $S_\ell$ independently from a
distribution $q\in\Delta_K$, the probability simplex on
$\{0,\ldots,K-1\}$, and plan to open from that phase onward
\[
H_{\ell K+r}=\ind\{r\ge S_\ell\} \, ,\qquad G_t=C_tH_t \, ,
\]
where $C_t$ is the availability indicator \eqref{eq:availability_indicator}.
Such delayed openings have been used practically to suppress strong early returns and
preserve sensitivity to weaker later photons~\cite{tosi2011fast,buttafava2015non},
but these works do not optimize the distribution of opening phases under
Fisher-information-based estimation criteria as we do here. Suffix plans require only one random start per period and keep
scheduled switching rare. This is desirable because gate edges have finite
rise time and can introduce timing jitter and excess noise~\cite{buttafava2014time,tosi2011fast}.

\subsubsection{Step 2. The induced gating frequencies}
For a start distribution $q$, write $\beta_r=\Pr(S_\ell\le r)=\sum_{s=0}^r q_s$. The induced frequencies then take the following form.
\begin{proposition}
\label{prop:suffix_gamma}
Let $p\in[0,1)^K$, $D\ge0$, and $\beta_r=\sum_{s\le r}q_s$.
Set $\mathsf B=A-I$ and $L_{rj}=p_j\ind\{1\le r-j\le D\}$ with non-cyclic difference $r-j$. The suffix frequencies are the unique solution of
\begin{equation}
\label{eq:suffix_gamma_matrix}
M\gamma=\beta,\qquad M=I+L+\Diag(\beta)(\mathsf B-L)\,.
\end{equation}
\end{proposition}
\begin{proof}
The frequencies exist by \Cref{lem:exogenous_gating_balance}. Multiply $C_t=1-\sum_{u=1}^D Y_{t-u}$ by $H_t$, take expectations, and average over periods. A detection earlier in the current period implies $H_t=1$. Every earlier-period detection is independent of the current suffix start. In the limit, these contributions are $L\gamma$ and $\Diag(\beta)(\mathsf B-L)\gamma$, giving \eqref{eq:suffix_gamma_matrix}. Uniqueness follows from \Cref{lem:gating_matrix_nonsingular}.
\end{proof}

Write $\gamma(\theta;q)$ for these suffix frequencies evaluated at $p=p(\theta)$.

\subsubsection{Steps 3--4. Choosing and optimizing the design objective}
To compare and optimize designs, we use the per-period range-variance coefficient
[equivalent to $A$-optimality with a weight matrix $W=\Diag(0,1,0)$]
\[
L_R^{\mathrm{period}}:=(c/2)^2[\mathcal I_{\mathrm{period}}^{-1}]_{\tau\tau}\,,
\]
where $\mathcal I_{\mathrm{period}}=K\mathcal I$ (\eqref{eq:Irate_def}) and the
factor $(c/2)^2$ accounts for the transformation $R=c\tau/2$~\cite[Chs.~3 and~5]{van2000asymptotic}.
For $N$ laser periods,
the efficient asymptotic range-variance benchmark is $L_R^{\mathrm{period}}/N$.
We choose a single start law $q$ to minimize the worst-case coefficient,
\begin{equation}
\label{eq:suffix_lidar_minimax}
\inf_{q\in\Delta_K}\max_{R\in[2,5]\,\mathrm m}
L_R^{\mathrm{period}}(\theta_R;\gamma(\theta_R;q))\,.
\end{equation}
This is a worst-case weighted $A$-optimality criterion, analogous to the unweighted objective in \eqref{eq:q_ranging_objectives}. Depending on the application, alternatives such as the weighted supremum in \Cref{rem:relative_objective} may also be appropriate. This and similar gating optimization problems are typically non-convex, but local optimization still allows one to improve over any given starting design and lower-bound the attainable improvement.

For each trial $q$ and range sample, we compute the gating frequencies
from \eqref{eq:suffix_gamma_matrix} and evaluate the objective. We optimize $q$ using the SciPy~\cite{virtanen2020scipy} implementation of L-BFGS-B~\cite{LBFGSB} from four fixed initial distributions (uniform, free-running, two discretized Gaussians), retaining the best result. We approximate the worst case by log-sum-exp; replacing this with an exact epigraph formulation changes the evaluated worst-case loss by less than $0.04\%$ in both regimes below. Since the problem is nonconvex,
global optimality is not guaranteed and the achieved reduction in worst-case
loss gives a lower bound on the improvement possible within the suffix
family over the sampled ranges.  (Our best local minima came from uniform initialization for low pile-up and from Gaussian for high pile-up, though these initializations produced nearly the same results regardless; in the high-pile-up regime, the free-running initialization yielded a loss approximately $2\times$ worse than the best result.)

\begin{figure}[htbp]
  \centering
  \includegraphics[width=\linewidth]{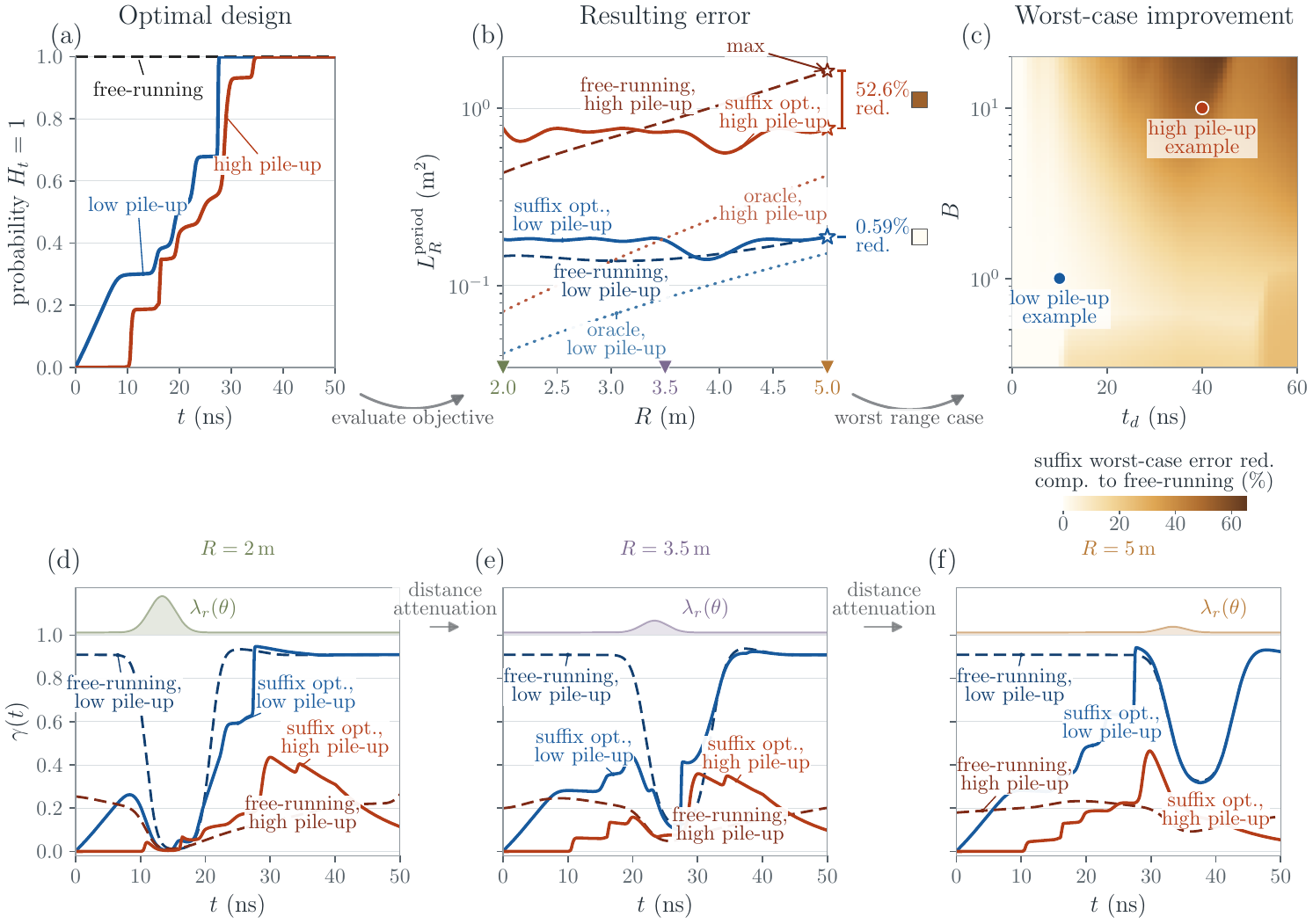}
  \caption{Minimax suffix-randomized gating for distance-attenuated lidar. (a) Optimized planned-open
  probabilities. (b) Per-period range-variance coefficient
  $L_R^{\mathrm{period}}$.
  (c) Relative reduction in worst-case range loss as a function of \(B\) and \(t_d\). (d)--(f) Realized gating frequencies at three
  ranges, with the corresponding attenuated pulses shown above. In panels (a) and (d)--(f), $t$ denotes phase time in nanoseconds; $\gamma(t)$ represents the phasewise gating frequencies $\gamma_r(\theta_R)$. Solid, dashed, and dotted curves show optimized suffix gating, free-running, and the pointwise oracle, respectively.
  }
  \label{fig:navspdp_suffix_composite}
\end{figure}

The examples in \Cref{fig:navspdp_suffix_composite} use the period and bin
size of \Cref{sec:exp_fr_regret}, and use the normalized wrapped-Gaussian pulse of
\Cref{subsection:modeling_single_photon_detection} with \(\sigma=2\,\mathrm{ns}\),
 \(R_\star=7.5\,\mathrm{m}\), and \(S_\star=0.5\). We compare two regimes:
\[
\begin{aligned}
&\text{low pile-up (lower background and dead time):} && B=1,\quad t_d=10\,\mathrm{ns}\,,\\
&\text{high pile-up (higher background and dead time):} && B=10,\quad t_d=40\,\mathrm{ns}\,.
\end{aligned}
\]
The optimized suffix rule reduces the worst-case loss by $0.59\%$ in the
lower-background, shorter-dead-time regime and by $52.6\%$ in the higher
regime, relative to free-running. The numerically optimized fixed-$\alpha$
designs improve on suffix gating by less than $10\%$ of the free-running
loss in both regimes.
In the higher regime, delaying the planned opening (panel (a)) reduces
dead time before the return, leading to concentration of detector availability
near the return locations (panels (d)--(f)). Panel (c) shows how the gain
varies with background and dead time. It need not increase monotonically
with dead time, since both the optimized loss and the free-running baseline
change.

To validate the Fisher-information prediction for the optimized suffix design at realistic acquisition times, we simulated $5{,}000$
independent trajectories of $N=10^5$ periods ($10\,\mathrm{ms}$) at $R\in\{2,3.5,5\}\,\mathrm m$ in both pile-up regimes, estimating $(a,\tau,b)$ with the robust pilot followed by a one-step correction introduced in~\citeasnoun{jorgensen2026fundamental}. In all cases, the empirical range MSE agreed with $L_R^{\mathrm{period}}/N$ to within $1.5$ Monte Carlo standard errors, estimated from the sample variance of the squared range errors.

\begin{remark}
\label{rem:relative_objective}
For relative range accuracy, one can replace the objective in
\eqref{eq:suffix_lidar_minimax} by its relative version $L_R^{\mathrm{period}}/R^2$. 
With the same model and range interval, numerically re-optimizing the suffix law for this criterion reduces the worst-case relative variance coefficient by $53.4\%$ and $53.7\%$ compared with free-running in the low- and high-pile-up regimes, respectively.
\end{remark}

\begin{remark}
To test whether attenuation can further improve the suffix designs, we also optimized the optical transmission jointly with the start
distribution. Joint optimization with gating selected full transmission in both regimes,
whereas optimizing attenuation without gating reduced the worst-case free-running loss by $7.6\%$
in the higher regime (\Cref{app:attenuation}).
\end{remark}

\section{Discussion}

Our results give practical pile-up dependent criteria for deciding when gating can lead to a substantial benefit over  free-running. When the conditions for free-running optimality fail,  we show that substantial gains are possible using a practical gating method for the case of distance-attenuated lidar. To demonstrate these gains in practice, we are currently investigating experimental implementations of similar lidar gating designs. Beyond lidar, the same design approach could benefit other applications such as fluorescence lifetime measurements~\cite{isbaner2016dead} and estimation of tissue optical
properties from time-resolved diffuse optical data~\cite{tosi2011fast}. Further directions include extending the framework to paralyzable detectors~\cite{muller1973dead} and additional noise mechanisms such as afterpulsing~\cite{cova1996avalanche}, as well as developing gating designs robust to model misspecification~\cite{white1982maximum}. A continuous-time formulation could connect optimal gating to statistical inference for Poisson processes~\cite{kutoyants2012statistical}. Extending the model to spatially coupled parameters or detector arrays could also provide a theoretical basis for global multipixel reconstruction methods that exploit spatial structure~\cite{kirmani2014first,vicente2013improved,lualdi2026quantum,lualdi2025fast,shin2015photon,liu2019pile,melidonis2023efficient}. A complementary engineering opportunity is to identify the most effective hardware architectures for gating, both to simplify implementation and to help close the remaining gap between practical schemes and the oracle gating.

\appendix

\section{Proofs for the optimal-gating problem}
\label{app:proofs_gating}

\subsection{Gating frequencies and dead-time constraints}
\label{app:controller_balance}
Write $(T_Dx)_r=\sum_{u=1}^{D}x_{r-u}$, with cyclic indices, so
$A(p)=I+T_D\operatorname{diag}(p)$. Repeated phases are counted
with their multiplicities when $D\ge K$.

\begin{lemma}
\label{lem:gating_matrix_nonsingular}
Let $p\in[0,1)^K$, $\beta\in[0,1]^K$, and $D\ge0$. Define
\[
B:=T_D\operatorname{diag}(p),\qquad
L_{rj}=p_j\ind\{1\le r-j\le D\}\,,
\]
using ordinary differences in $L$.
Then $M=I+L+\Diag(\beta)(B-L)$ is nonsingular.
In particular, $A(p)=I+B$ is nonsingular.
\end{lemma}
\begin{proof}
Suppose $Mv=0$. Define $F_j=\sum_{i=0}^{j-1}p_iv_i$ for
$0\le j\le K$, put $C=F_K$, and extend by $F_{j+K}=F_j+C$.
In this notation, $Mv = 0$ implies
\[
F_{r+1}=(1-p_r)F_r+\kappa_rp_rF_{r-D},\qquad
\kappa_r=\begin{cases}1,&r\ge D\,,\\ \beta_r,&r<D\,.\end{cases}
\]
Choose the sign of $v$ so that $C\ge0$. Suppose
$m=\max_{0\le j\le K}F_j>0$, then $F_{r-D}\le m$ for all
$0\le r<K$. At the first occurrence of $m$, its predecessor is
smaller, so the recursion gives a value strictly below $m$, since
$1-p_r>0$ and $0\le\kappa_r\le1$. This is a contradiction.
Hence $C=0$. Applying the same argument to $F$ and $-F$ gives
$F=0$, and the original equations give $v=0$.
Taking $\beta=\mathbf1$ yields $M=A(p)$.
\end{proof}

\begin{lemma}
\label{lem:controller_balance}
For every $p\in[0,1)^K$, $A(p)$ is nonsingular and
$s(p):=A(p)^{-1}\mathbf1$ is strictly positive.
For $\alpha\in[0,1]^K$, put $p^\alpha=\alpha\odot p$.
The $\alpha$-scheme has limiting gating frequencies
\begin{equation}
\label{eq:controller_frequency_formula}
\Gamma(p,\alpha)=\operatorname{diag}(\alpha)A(p^\alpha)^{-1}\mathbf1\,.
\end{equation}
This map is jointly smooth and locally Lipschitz in $p$ uniformly over $\alpha$.

Let $\mathcal H_{t_0}$ be a sigma-algebra describing information available
at a period boundary $t_0$, and let $R_{t_0}\in\{0,\ldots,D\}$ and
$\widehat\alpha\in[0,1]^K$ be $\mathcal H_{t_0}$-measurable.
Run the scheme for $m$ periods with opening probabilities fixed at
$\widehat\alpha$, with the block's photon variables and gate draws independent
of $\mathcal H_{t_0}$. For $g_r=m^{-1}\sum_{\ell=0}^{m-1}G_{t_0+r+\ell K}$,
\begin{equation}
\label{eq:block_uniform_bound}
\mathbb E_p\!\left[\|g-\Gamma(p,\widehat\alpha)\|_2
\,\middle|\,\mathcal H_{t_0}\right]
\le C_p m^{-1/2}\qquad\text{almost surely}\,,
\end{equation}
where $C_p$ depends only on $p,K,D$.
The unconditional bound follows by taking expectations.
\end{lemma}
\begin{proof}
Nonsingularity of $A(p^\alpha)$ follows from \Cref{lem:gating_matrix_nonsingular} with $\beta=\mathbf1$.

For a block of $m$ periods, let $c,g,y\in[0,1]^K$ denote the vectors of
phasewise empirical averages of $C_t,G_t,Y_t$, respectively.
Use the filtration
\[
\mathcal F_t=\mathcal H_{t_0}\vee\sigma\bigl(U_s,Y_s:t_0\le s<t\bigr),
\qquad t\ge t_0\,,
\]
where $U_s$ are the independent uniform draws in \Cref{alg:alpha_scheme}.
For $r=t\bmod K$, conditional on $\mathcal F_t$, $G_t$ and $Y_t$ have
expectations $\widehat\alpha_rC_t$ and $p_r^{\widehat\alpha}C_t$, respectively. Hence $G_t-\widehat\alpha_rC_t$
and $Y_t-p_r^{\widehat\alpha}C_t$ are bounded martingale differences. Their phase
averages $e,z$ therefore give
\[
\begin{gathered}
g=\operatorname{diag}(\widehat\alpha)c+e,\qquad
y=\operatorname{diag}(p^{\widehat\alpha})c+z\,,\\
\mathbb E_p[e_r^2\mid\mathcal H_{t_0}],\quad
\mathbb E_p[z_r^2\mid\mathcal H_{t_0}]\le m^{-1}\,.
\end{gathered}
\]
The dead-time recursion~\eqref{eq:dead_time_counter} gives
$c=\mathbf1-T_Dy+b$ for a boundary correction $b\in\mathbb R^K$.
If $H_r(h)$ counts phase-$r$ bins among the next $h$
bins at a period boundary, then
$b_r=[H_r(R_{\rm out})-H_r(R_{\rm in})]/m$,
where $R_{\rm in}:=R_{t_0}$ and $R_{\rm out}:=R_{t_0+mK}$ are the
remaining dead-time counters at the start and end of the block.
Since $R_{\rm in},R_{\rm out}\in\{0,\ldots,D\}$, both counts lie in
$[0,\lceil D/K\rceil]$, so $\|b\|_\infty\le\lceil D/K\rceil/m$.
Substituting the three preceding identities into one another:
\[
g-\operatorname{diag}(\widehat\alpha)s(p^{\widehat\alpha})
=e+\operatorname{diag}(\widehat\alpha)A(p^{\widehat\alpha})^{-1}(b-T_Dz)\,.
\]
The inverse is uniformly bounded on the compact rectangle
$0\le p^{\widehat\alpha}\le p<\mathbf1$, proving
\eqref{eq:block_uniform_bound} by the conditional second-moment bounds above.
Taking expectations gives the unconditional bound and, for fixed opening
probabilities, the frequency formula. For free running, no incident event in the preceding $D$ bins
guarantees availability. Apply the block bound with $\alpha=\mathbf1$
and let $m\to\infty$ in expectations, ignoring the first $D$ bins, to obtain
\begin{equation}\label{eq:fr_availability_lower}
1\ge s_r(p)\ge\prod_{u=1}^D(1-p_{r-u})>0\,.
\end{equation}
Finally, the inverse formula gives smoothness and uniform local
Lipschitz continuity.
\end{proof}

\subsection[Proof of the polytope characterization]
{Proof of \Cref{thm:gating_constraints}}
\label{app:proof_gating_constraints}
\begin{proof}
For necessity, the causal differences
$\xi_t=G_t(Z_t-p_{t\bmod K})$ are bounded martingale
differences when conditioned after choosing $G_t$ and before
observing $Z_t$. Their phase averages therefore have second
moment at most $1/L$. Sum the inequality
\begin{equation}
\label{eq:basic_ineq}
G_t+\sum_{u=1}^{D}Y_{t-u}\le1
\end{equation}
over phase $r$, substitute $Y_t=p_{t\bmod K}G_t+\xi_t$,
and let $L\to\infty$, which gives  $A(p)\gamma\le\mathbf1$.

Conversely, suppose $\gamma\ge0$ and $A(p)\gamma\le\mathbf1$,
and set $s=\mathbf1-T_D\operatorname{diag}(p)\gamma$.
Then $0\le\gamma\le s$, and every $s_r$ is positive, since
\[
s_{r+1}
=s_r-p_r\gamma_r+p_{r-D}\gamma_{r-D}
\ge(1-p_r)s_r
\]
and a zero would hence imply $s=\gamma=0$, contrary to the definition of $s$.
Set $\alpha_r=\gamma_r/s_r$. Then
\[
A(\alpha\odot p)s=\mathbf1,\qquad
\Gamma(p,\alpha)=\operatorname{diag}(\alpha)A(\alpha\odot p)^{-1}\mathbf1
=\gamma\,.
\]
\Cref{lem:controller_balance} proves phase convergence. The other DED
conditions follow directly from the causal gate rule and
counter update.
\end{proof}

\subsection[Proof of the free-running corollary]
{Proof of \Cref{cor:free-running}}
\label{app:proof_free_running}
\begin{proof}
The free-running scheme satisfies
$$
G_t+\sum_{u=1}^{D}Y_{t-u}= 1,\qquad t\ge 0\,,
$$
so the same argument leading to \eqref{eq:basic_ineq} applies with equality and yields $A\gamma^{\mathrm{fr}}=\mathbf{1}$.
\end{proof}

\subsection[Proof of adaptive oracle attainment]
{Proof of \Cref{thm:adaptive_oracle}}
\label{app:proof_adaptive_oracle}
We specify the regularized oracle construction and then prove its
convergence.

For $\eta\in\tilde U$, denote the feasible set and objective of
\eqref{eq:gating_convex_problem} at $\theta_0=\eta$ by
\[
\mathcal P(\eta):=
\{\gamma\ge0:A(\eta)\gamma\le\mathbf1\},
\qquad
f_\eta(\gamma):=
\Tr\!\bigl(\mathcal I(\eta;\gamma)^{-1}\bigr)\,.
\]
Define the unique
minimum-norm oracle by
\[
\gamma^\dagger(\eta):=
\arg\min\left\{\|\gamma\|_2^2:
\gamma\in\arg\min_{v\in\mathcal P(\eta)}f_\eta(v)\right\}\,.
\]
The oracle-optimal set is nonempty, compact, and convex, so this
minimum-norm element exists and is unique.

The construction applies to any pilot satisfying
$\sqrt m(\hat\theta_m-\theta)=O_{P_\theta}(1)$ for every
$\theta\in\tilde U$. In \Cref{alg:adaptive_plugin}, the
$L_{\rm pil}-K\ge D$ closed bins before each probe ensure availability.
Thus the pilot pairs $(J_m,X_m)$ are iid, with $J_m$ uniform and
$X_m\mid J_m=r\sim\Ber(p_r(\theta))$. The finite observation space,
identifiability using all phases, and \Cref{ass:lan} give consistency
and the stated rate for the pilot MLE
\cite[Theorems~5.7 and~5.39]{van2000asymptotic}.
In \Cref{alg:adaptive_plugin}, choose the opening map at iteration $m$
as follows. Set
$\varepsilon_m=m^{-1/4}$. When $\hat\theta_m\in\tilde U$, compute
\begin{equation}
\label{eq:adaptive_regularized_oracle}
\hat\gamma_m\in\arg\min_{\gamma\in\mathcal P(\hat\theta_m)}
\left\{f_{\hat\theta_m}(\gamma)
+\varepsilon_m\|\gamma\|_2^2\right\}\,,
\end{equation}
and set
\[
(\alpha_m)_r=
\frac{(\hat\gamma_m)_r}
{1-\sum_{u=1}^D p_{r-u}(\hat\theta_m)(\hat\gamma_m)_{r-u}},
\qquad r=0,\ldots,K-1\,,
\]
whose denominators are strictly positive. Set $\alpha_m=0$ when
$\hat\theta_m\notin\tilde U$.
We show that the realized phasewise gating frequencies converge in
$L^1(P_\theta)$ to $\gamma^\dagger(\theta)$.

\begin{proof}
Fix $\theta\in\tilde U$ and write $\gamma^\dagger=\gamma^\dagger(\theta)$
and $\Gamma(\theta,\alpha):=\Gamma(p(\theta),\alpha)$.
The pilot-rate hypothesis gives
$\delta_m:=\|\hat\theta_m-\theta\|_2=o_{P_\theta}(\varepsilon_m)$.
The pilot MLE has this rate as noted above.
Choose opening probabilities $\alpha^\dagger$ realizing $\gamma^\dagger$
at $\theta$, and put $x_m=\Gamma(\theta,\alpha_m)$.
At the estimated parameter,
$\Gamma(\hat\theta_m,\alpha^\dagger)$ is feasible for the regularized
problem, while $\Gamma(\hat\theta_m,\alpha_m)=\hat\gamma_m$.
The information matrix of this feasible design stays positive definite near
$\theta$, so comparison with its bounded regularized objective value bounds the optimizer's
inverse-information trace and keeps the smallest information
eigenvalue away from zero. \Cref{lem:controller_balance} and smoothness of
the information map therefore give the following regularized optimality bound
at the true parameter
\begin{equation}
\label{eq:regularized_oracle_comparison}
f_\theta(x_m)+\varepsilon_m\|x_m\|_2^2
\le f_\theta(\gamma^\dagger)
+\varepsilon_m\|\gamma^\dagger\|_2^2+C\delta_m
\end{equation}
on an event whose probability tends to one.
Since $x_m$ is feasible at $\theta$ and
$\delta_m/\varepsilon_m\to0$, it is asymptotically optimal and
$\|x_m\|_2^2\le\|\gamma^\dagger\|_2^2+o_P(1)$.
Compactness and uniqueness of the minimum-norm oracle give
$x_m\to\gamma^\dagger$ in probability and, by boundedness, in
$L^1$.

Let $g_m$ be the empirical phase frequencies in the $m$-period block
following the $m$th probe. The conditional bound in
\Cref{lem:controller_balance} applies because $\alpha_m$ is chosen before
that block and held fixed within it. Taking expectations gives
\[
\mathbb E_\theta\|g_m-\gamma^\dagger\|_2
\le C_{p(\theta)}m^{-1/2}
+\mathbb E_\theta\|\Gamma(\theta,\alpha_m)-\gamma^\dagger\|_2
\longrightarrow0\,.
\]
After $M$ iterations, the empirical frequencies over these blocks are
$\sum_{m=1}^M m g_m/\sum_{m=1}^M m$.
They converge in $L^1$ to $\gamma^\dagger$ because the errors above tend
to zero and every fixed collection of early blocks has vanishing weight.
The pilot intervals occupy $M L_{\rm pil}$ bins, whereas the other
blocks occupy $K M(M+1)/2$ bins, so the pilots do not affect the limit.
An unfinished iteration contributes only $O(M)$ bins and likewise does
not affect it. Hence the required phase limits hold at every horizon.
The incident-event model, causality, and dead-time constraint follow from
the construction. Together with the phase limits for every $\theta\in\tilde U$,
these establish \Cref{def:DED} for the family restricted to $\tilde U$.
\end{proof}
More generally, the same conclusion holds for any
$\varepsilon_m\downarrow0$ satisfying
$\|\hat\theta_m-\theta\|_2=o_{P_\theta}(\varepsilon_m)$.

\begin{proof}[Proof of \Cref{rem:adaptive_target}]
Fix $\theta\in\tilde U$. The opening map $\alpha$ of the continuous
feasible target $\gamma$ is continuous. A consistent pilot therefore gives
$\alpha_m\to\alpha(\theta)$ in probability, and
\eqref{eq:controller_frequency_formula} yields
\[
\Gamma(p(\theta),\alpha_m)\longrightarrow
\Gamma(p(\theta),\alpha(\theta))=\gamma(\theta)
\quad\text{in probability and in }L^1\,,
\]
where boundedness gives $L^1$ convergence. The block-averaging argument
above then gives the claimed phase limits.
\end{proof}

\section{Proofs for the free-running optimality results}
\label{app:proofs_fr_opt}

\subsection[Proof of the exact optimality certificate]{Proof of \Cref{prop:general_fr_optimal}}
\label{app:proof_general_fr_optimal}
\begin{proof}
Write $g=\gamma^{\mathrm{fr}}$. By \Cref{lem:controller_balance}, $g>0$, so
\eqref{eq:positivity_condition} gives $\mathcal I^{\mathrm{fr}}\succ0$.
For a feasible $\gamma$, put $z=\mathbf1-A\gamma\ge0$.
Then $h^\top(g-\gamma)=\nu^\top z$.
Convexity and $\nabla f(g)=-h$ make $g$ optimal exactly when this
quantity is nonnegative for every feasible $z$.
Since $g>0$, every sufficiently small $z\ge0$ is feasible, giving
the exact condition $\nu\ge0$.

For $N\ge0$ with $I+N$ nonsingular, we also have
\begin{equation}\label{eq:positive_box_inverse}
b\ge0,\quad Nb\le b
\quad\Longrightarrow\quad (I+N)^{-1}b\ge0\,.
\end{equation}
The map $v\mapsto b-Nv$ takes $[0,b]$ into itself. Brouwer's fixed-point theorem~\cite[Chapter~6]{border1985fixed} gives a fixed point, necessarily $(I+N)^{-1}b$. Apply this with $N=A^\top-I$ and
$b=h$ to obtain the local sufficient condition.
\end{proof}

\subsection[Proof of the regret bound]{Proof of \Cref{thm:regret_via_lambda_minus}}
\label{app:proof_regret}
\begin{proof}
Write $g=\gamma^{\mathrm{fr}}$. Since $0\le\gamma_r\le1$ and $g_r\ge m$, we have
$\mathcal I^{\mathrm{fr}}\succeq m\mathcal I(\theta;\gamma)$.
Inversion and taking the trace give the matrix bound and the first
scalar inequality. The product bound \eqref{eq:fr_availability_lower}
gives $m\ge e^{-\rho}$, proving the remaining inequality.
\end{proof}

\subsubsection{Sharpness from a two-bin model}
\label{app:cyclic_threshold_sharp}
The following example shows that the regret bound~\eqref{eq:fr_regret_sharp}
and the free-running optimality threshold $\rho_\star\le\log2$ in \Cref{thm:delay_optimality} are both sharp.
Take $K=2$, $D=1$, $p=(p_0,p_1)$ with $0<p_1<p_0<1$, and scalar
per-phase information $J_0=0$, $J_1=1$. The $\alpha$-scheme of
\Cref{alg:alpha_scheme}, with $(\alpha_0,\alpha_1)\in[0,1]^2$, has
frequencies satisfying $\gamma_0=\alpha_0(1-p_1\gamma_1)$ and
$\gamma_1=\alpha_1(1-p_0\gamma_0)$. Thus
\[
\gamma_1(\alpha_0,\alpha_1)
=\frac{\alpha_1(1-p_0\alpha_0)}{1-p_0p_1\alpha_0\alpha_1},
\qquad \mathcal I=\frac{\gamma_1(\alpha_0,\alpha_1)}2\,.
\]

When the informative phase is known, closing the uninformative bin gives
$\gamma_1(0,1)=1$, which is optimal since every feasible $\gamma_1\le1$.
Hence $f_\star=2$ and $m=\min_r\gamma_r^{\mathrm{fr}}=\gamma_1(1,1)$, so
\[
\frac{f(\gamma^{\mathrm{fr}})-f_\star}{f(\gamma^{\mathrm{fr}})}
=1-m=\frac{p_0(1-p_1)}{1-p_0p_1}
\longrightarrow p_0=1-e^{-\rho}
\qquad(p_1\downarrow0)\,,
\]
with $p_0$ fixed and $\rho=-\log(1-p_0)$. This attains the first bound
in~\eqref{eq:fr_regret_sharp} and approaches the second.

For an unknown cyclic shift, use the same opening probability
$\alpha\in[0,1]$ at both phases, with independent planned gates.
This is a stationary ergodic pre-planned schedule, and its information
under either shift is $\gamma_1(\alpha,\alpha)/2$. Moreover,
\[
\gamma_1(\alpha,\alpha)>\gamma_1(1,1)
\quad\Longleftrightarrow\quad
\frac{1-p_0}{p_0(1-p_1)}<\alpha<1\,.
\]
This interval is nonempty exactly when $p_0>1/(2-p_1)$, whose threshold
tends to $1/2$ as $p_1\downarrow0$. Thus examples improving both the
uniform-average and worst-case inverse-information loss have pile-up
arbitrarily close to $\log2$ from above.

\subsection[Pre-planned schedules and free-running optimality]{Pre-planned schedules and proof of \Cref{thm:delay_optimality}}
\label{app:block_randomized}

\begin{lemma}
\label{lem:exogenous_gating_balance}
Let $\mathsf Q$ be the stationary ergodic law of a pre-planned schedule
as in \Cref{subsubsection:preplanned_schedules}, and let $p\in[0,1)^K$.
The empirical averages in \eqref{eq:phase_convergent} converge to the
gating frequencies $\gamma_r(\mathsf Q;p)$ almost surely under the joint
law of the planned gates $(H_t)$ and photon events $(Z_t)$.
With $w_r=\Pr(H_r=0)$,
\begin{equation}\label{eq:exogenous_balance_bounds}
w_r\prod_{u=1}^D(1-p_{r-u})
\le[\mathbf1-A(p)\gamma]_r\le w_r\,.
\end{equation}
\end{lemma}
\begin{proof}
For $D=0$, $G_t=H_t$ and the result is immediate. Assume $D\ge1$.
The paired period vectors of planned gates and photon events are stationary
and ergodic, since the photon vectors are iid and independent of the
plan~\cite[Chapter~2]{einsiedler2011ergodic}. 

For $t\ge mK\ge D$, define the approximation
\[
G_t^{(m)}:=H_t\ind\{R_{t;t-mK}=0\}\,,
\]
where $R_{u;s}$ follows~\eqref{eq:dead_time_counter} from $R_{s;s}=0$,
with $Y_u$ replaced by $H_uZ_u\ind\{R_{u;s}=0\}$.
A run of $D$ photon-free bins makes both counters zero and their later
gates identical. By independence of disjoint photon blocks, the
probability of no such reset is at most
\[
\varepsilon_m=[1-(1-\max_r p_r)^D]^{\lfloor mK/D\rfloor}\longrightarrow0\,.
\]
The approximations and reset-failure indicators are finite-window functions
of the stationary ergodic input. Birkhoff's theorem~\cite[Theorem~2.30]{einsiedler2011ergodic} gives deterministic
phase-average limits, respectively $a_{m,r}$ and at most $\varepsilon_m$.
Gate disagreement requires reset failure, so, almost surely,
\[
\limsup_{n\to\infty}
\left|\frac1n\sum_{\ell=0}^{n-1}G_{\ell K+r}-a_{m,r}\right|
\le\varepsilon_m\,.
\]
As $\varepsilon_m\to0$, the phase average converges to a deterministic
limit $\gamma_r=\lim_{m\to\infty}a_{m,r}$.

The unused availability $C_t(1-H_t)$ is at most $1-H_t$, with equality
after $D$ photon-free bins. Thus, for $t\ge D$,
\[
(1-H_t)\prod_{u=1}^D(1-Z_{t-u})
\le C_t(1-H_t)=1-G_t-\sum_{u=1}^D Y_{t-u}\le1-H_t\,.
\]
Taking expectations and phase averages gives~\eqref{eq:exogenous_balance_bounds}.
\end{proof}

\begin{proposition}
\label{lem:q_gamma_exists}
Let $p\in[0,1)^K$, $D\ge0$, and consider a pre-planned schedule
as in \Cref{subsubsection:preplanned_schedules}, whose period vectors
$H^{(\ell)}$ are iid with law $Q$ on $\{0,1\}^K$. With $e_j$ the
coordinate vectors in $\mathbb R^{D+1}$, define the counter transitions
\[
\mathsf M_r(h)_{j,\cdot}=
\begin{cases}
(1-hp_r)e_0^\top+hp_re_D^\top,&j=0\,,\\
e_{j-1}^\top,&1\le j\le D\,.
\end{cases}
\]
The period transition matrix
\[
\mathsf P_Q=\mathbb E_{H\sim Q}
\bigl[\mathsf M_0(H_0)\cdots\mathsf M_{K-1}(H_{K-1})\bigr]
\]
has a unique stationary row distribution $\mu_Q$ satisfying
$\mu_Q\mathsf P_Q=\mu_Q$ and $\mu_Q\mathbf1=1$.
The gating frequencies are
\[
\boxed{\gamma_r(Q;p)=\mathbb E_{H\sim Q}\!
\left[H_r\bigl(\mu_Q\mathsf M_0(H_0)\cdots
\mathsf M_{r-1}(H_{r-1})\bigr)_0\right]\,,}
\]
with an empty product equal to the identity.
\end{proposition}
\begin{proof}
The period-boundary counter is a finite Markov chain with transition
matrix $\mathsf P_Q$, started at zero. A sufficiently long photon-free
block sends every counter state to zero with uniformly positive
probability. Hence the chain has a unique stationary distribution
$\mu_Q$, and its distribution converges to $\mu_Q$ by coupling~\cite[Chapter~5]{levin2017markov}.
The fresh planned block $H$ is independent of its initial counter.
Conditioning on $H$ and propagating the counter through the phase
transitions gives the displayed expression for the limiting opening
probability. Averaging over periods and using bounded convergence
in \Cref{lem:exogenous_gating_balance} identifies it with $\gamma_r$.
\end{proof}

\begin{theorem}
\label{thm:q_extension_free_running}
Fix $p\in[0,1)^K$ and $J_r\in\mathbb S_+^d$ with
$\sum_rJ_r\succ0$. For $j\in\mathbb Z_K$, let
$p^{(j)}_r=p_{r-j}$ and $J^{(j)}_r=J_{r-j}$.
Let $\pi$ be a prior on these shifts, with
$\pi_{\min}=\min_j\pi_j$ and $\pi_{\max}=\max_j\pi_j$. If
\begin{equation}\label{eq:cyclic_prior_condition}
\rho\le\log\!\left(1+\frac{\pi_{\min}}{\pi_{\max}}\right)\,,
\end{equation}
then every pre-planned schedule $\mathsf Q$, with shifted information
$F_j:=K^{-1}\sum_r\gamma_r(\mathsf Q;p^{(j)})J^{(j)}_r$, satisfies
\begin{equation}\label{eq:cyclic_information_covariance}
\sum_j\pi_jF_j\preceq \mathcal I^{\mathrm{fr}}\,,
\end{equation}
where $\mathcal I^{\mathrm{fr}}\succ0$ is the common free-running information.
\end{theorem}
\begin{proof}
Put $A=A(p)$ and $\sigma_t=\prod_{u=1}^D(1-p_{t-u})$.
First, $A^{-1}(\mathbf1-z)\ge0$ whenever $0\le z\le\sigma$.
Indeed, the $\alpha$-scheme has continuous availability
$s(\alpha)=A(\alpha\odot p)^{-1}\mathbf1\ge\sigma$ by the no-photon event.
Brouwer's fixed-point theorem~\cite[Chapter~6]{border1985fixed} gives a fixed point of the continuous cube map
$\alpha\mapsto\mathbf1-z\oslash s(\alpha)$, with componentwise division.
At that point, its frequencies $\gamma=\alpha\odot s(\alpha)$ satisfy
$\mathbf1-A\gamma=(\mathbf1-\alpha)\odot s(\alpha)=z$, proving the claim.

Let $\bar\gamma_t=\sum_j\pi_j\gamma_{t+j}(\mathsf Q;p^{(j)})$.
Applying \eqref{eq:exogenous_balance_bounds} to each cyclic shift and
averaging gives, with
$S=\sum_r\Pr(H_r=0)$ and $\bar\eta=\mathbf1-A\bar\gamma$,
\[
\pi_{\min}S\sigma\le\bar\eta\le\pi_{\max}S\mathbf1\,.
\]
If $S=0$, then $\bar\eta=0$. Otherwise,
$z:=\mathbf1-\bar\eta/(\pi_{\max}S)$ satisfies
$0\le z\le\mathbf1-(\pi_{\min}/\pi_{\max})\sigma\le\sigma$
by \eqref{eq:cyclic_prior_condition}. The claim therefore gives
$\gamma^{\mathrm{fr}}-\bar\gamma=A^{-1}\bar\eta\ge0$.
Uniqueness of the free-running balance solution gives cyclic equivariance.
The positive free-running frequencies give $\mathcal I^{\mathrm{fr}}\succ0$. Thus
\[
\sum_j\pi_jF_j=\frac1K\sum_t\bar\gamma_tJ_t
\preceq\frac1K\sum_t\gamma_t^{\mathrm{fr}}J_t=\mathcal I^{\mathrm{fr}}\,.
\]
\end{proof}

For a uniform cyclic prior, \eqref{eq:cyclic_prior_condition} is precisely
$\rho\le\log2$, including equality. For a strictly positive
nonuniform prior and $\rho>0$, condition
\eqref{eq:cyclic_prior_condition} is equivalently
$\pi_{\max}/\pi_{\min}\le1/(e^{\rho}-1)$.

\begin{proof}[Proof of \Cref{thm:delay_optimality}]
For a bounded prior density $\pi$, define
\[
s_\pi(\omega):=\sum_{j=0}^{K-1}\pi(\omega+j\Delta),
\qquad
w_j(\omega):=\frac{\pi(\omega+j\Delta)}{s_\pi(\omega)}
\]
where $s_\pi(\omega)>0$. For almost every such offset,
\[
\frac{\min_jw_j(\omega)}{\max_jw_j(\omega)}
\ge\frac{\pi_{\min}}{\pi_{\max}}\,.
\]
The prior-dependent pile-up condition therefore allows us to apply
\Cref{thm:q_extension_free_running} with $p=p(\omega)$,
$J_r=J_r(\omega)$, and weights $w_j=w_j(\omega)$.
Write $F_j=\mathcal I_{\omega+j\Delta}(\mathsf Q)$ and
$\mathcal I^{\mathrm{fr}}=\mathcal I_\omega(\mathsf Q^{\mathrm{fr}})$.
If the positively weighted $F_j$ are nonsingular, operator convexity and
order reversal of inversion~\cite[Section~3.6]{boyd2004convex} give
\[
\sum_{j:w_j>0}w_jF_j^{-1}
\succeq\left(\sum_jw_jF_j\right)^{-1}
\succeq (\mathcal I^{\mathrm{fr}})^{-1}\,.
\]
Taking a (weighted) trace gives the cyclic loss comparison. Integrating against $s_\pi(\omega)\,d\omega$ proves the
prior-averaged claim. Taking the supremum over offsets gives the worst-case result.
\end{proof}

\section{Joint attenuation and gating}
\label{app:attenuation}
Besides gating, another approach for more informative single-photon experiments is  attenuating the incoming light, trading off reduced pile-up against the loss of incident photons~\cite{gupta2019photon}. We consider joint design of attenuation with gating and  show that at low per-bin photon counts, attenuation can help when no gating is used, but offers little to no additional benefit once gating is freely optimized.

Throughout this appendix, we neglect dark counts, which are not reduced by optical attenuation. More formally, we assume a designable attenuation factor $u\in(0,1]$ scales both signal and optical background, i.e. $\lambda_r^u(\theta)=u\lambda_r(\theta)$.  Fixing $\theta$, write $\lambda_r=\lambda_r(\theta)>0$ and $g_r=\nabla\lambda_r(\theta)$, so that
\[
p_r(u)=1-e^{-u\lambda_r},\qquad
J_r(u)=j_r(u)g_rg_r^\top,\qquad
j_r(u)=\frac{u^2}{e^{u\lambda_r}-1}\,.
\]
As motivated in \Cref{section:gating_opt_problem}, we design this parameter together with the gating frequencies $\gamma$ that lie in the convex polytope $\mathcal P_u=\{\gamma\ge0:A(p(u))\gamma\le\mathbf1\}$, with information matrix
$\mathcal I_u(\gamma)=K^{-1}\sum_r\gamma_rJ_r(u)$. Define the oracle values as 
\[
f_\star(u):=\inf_{\gamma\in\mathcal P_u}\Tr(\mathcal I_u(\gamma)^{-1}),
\qquad f_\star^{\mathrm{joint}}:=\inf_{0<u\le1}f_\star(u)\,.
\]
\begin{proposition}
\label{prop:attenuation_gating}
Assume some feasible gating scheme gives an invertible information matrix. Let
$M=\max_r\lambda_r$ and let $\rho$ be the pile-up defined in
\eqref{eq:dead_time_photon_count} at full transmission ($u=1$). Define
\[
\chi(x)=\left(\frac{x}{2\sinh(x/2)}\right)^2,
\qquad \psi(x)=1+e^x-\frac{2(e^x-1)}x\,.
\]
The relative regret of optimal gating at full transmission, compared with jointly optimizing gating and attenuation, satisfies the bound
\begin{equation}\label{eq:attenuation_bound}
0\le\frac{f_\star(1)-f_\star^{\mathrm{joint}}}{f_\star(1)}
\le1-\chi(M)\,.
\end{equation}
Further, this regret is zero if $e^{-\rho}\ge\psi(M)$. In particular, this is the case if $D\ge1$ and $\rho\le\log2$.
\end{proposition}
\begin{proof}
For $\gamma\in\mathcal P_u$, set
$\bar\gamma_r=\gamma_rp_r(u)/p_r(1)$.
This preserves detection frequencies while reducing gating frequencies,
so $\bar\gamma\in\mathcal P_1$. Moreover,
\[
\frac{j_r(1)/p_r(1)}{j_r(u)/p_r(u)}
=\left(\frac{\sinh(u\lambda_r/2)}{u\sinh(\lambda_r/2)}\right)^2
\ge\chi(\lambda_r)\ge\chi(M)\,.
\]
Hence $\mathcal I_1(\bar\gamma)\succeq\chi(M)\mathcal I_u(\gamma)$ which gives  \eqref{eq:attenuation_bound}.

For zero regret, increase transmission from $u$ to $1$ along the information-preserving path
$\gamma_r(v)=\gamma_rj_r(u)/j_r(v)$, $u\le v\le1$, and write
$F_r(v)=(A(p(v))\gamma(v))_r$. At any feasible point with an active constraint $F_r(v)=1$, $\gamma_r(v)$ equals the availability fraction and hence is at least $e^{-v\rho}$, since no latent arrivals in the preceding $D$ bins guarantee availability. At such a point, differentiating and using that $x/(1-e^{-x})$ and $x\coth(x/2)$ are increasing gives
\[
vF_r'(v)\le\frac{vM}{e^{vM}-1}
\bigl[\psi(vM)-\gamma_r(v)\bigr]<0\qquad(v<1)\,,
\]
since $\psi$ increases and $\psi(vM)<\psi(M)\le e^{-\rho}\le e^{-v\rho}$. Thus every active constraint points strictly inward for $v<1$, preventing a first exit from feasibility; continuity gives feasibility at $v=1$.
If $D\ge1$ and $\rho\le\log2$, then indeed $M\le\rho$ and $\psi(M)\le\psi(\log2)<1/2\le e^{-\rho}$. 
\end{proof}
In both regimes of \Cref{subsec:lidar}, $M<0.151$ uniformly over the
considered range, so the relative regret is below $0.19\%$. This bound, however, 
allows optimization over the full polytope $\mathcal P_u$, whereas the
suffix schemes of \Cref{sec:practical_design_recipe} form a restricted
family. To assess whether attenuation helps within this family, we replace
$(a,\tau,b)$ by $(ua,\tau,ub)$ in \eqref{eq:suffix_gamma_matrix} and the
range-loss objective, and optimize $u$ jointly with the suffix start
distribution. In both regimes, the numerical optimization selected full
transmission ($u=1$), so allowing attenuation also provided no additional
reduction in the worst-case per-period range-variance coefficient.

For free-running acquisition, however, optimizing attenuation reduced this
coefficient by $7.6\%$ in the regime with higher background and longer dead
time, illustrating some benefit of attenuation when gating is unavailable.
\section*{Acknowledgments}

\ifarxiv
\paperfunding
\fi

ChatGPT was used for proofreading and minor language editing. OpenAI Codex
was used to assist with software development and plotting numerical results.
The authors reviewed all AI-assisted text and code and take full responsibility
for the manuscript, computations, and figures.

\bibliographystyle{siamplain}
\Needspace{6\baselineskip}
\bibliography{references}

\end{document}

%% file: ex_shared.tex
\usepackage{lipsum}
\usepackage{amsmath}
\usepackage{amsfonts}
\usepackage{graphicx}
\usepackage{tikz}
\usepackage{epstopdf}
\usepackage{algorithm}
\usepackage{cleveref}
\usepackage{algpseudocode}
\ifpdf
  \DeclareGraphicsExtensions{.eps,.pdf,.png,.jpg}
\else
  \DeclareGraphicsExtensions{.eps}
\fi

\usepackage{enumitem}
\setlist[enumerate]{leftmargin=.5in}
\setlist[itemize]{leftmargin=.5in}

\newcommand{\R}{\mathbb{R}}

\usepackage{dsfont}
\newcommand{\ind}{\mathds{1}}
\newcommand{\Ber}{\mathrm{Ber}}
\newcommand{\Tr}{\operatorname{tr}}
\newcommand{\Diag}{\operatorname{diag}}
\makeatletter
\newcommand{\citenumonly}[1]{\begingroup\def\@cite##1##2{##1\if@tempswa, ##2\fi}\cite{#1}\endgroup}
\makeatother
\newcommand{\citeasnoun}[1]{Ref.~\citenumonly{#1}}
\newcommand{\hide}[1]{}
\newcommand{\fj}[1]{}

\newsiamremark{remark}{Remark}
\newsiamremark{hypothesis}{Hypothesis}
\newsiamremark{assumption}{Assumption}

\crefname{hypothesis}{Hypothesis}{Hypotheses}
\crefname{assumption}{Assumption}{Assumptions}
\newsiamthm{claim}{Claim}
\newsiamremark{fact}{Fact}
\crefname{fact}{Fact}{Facts}

\headers{Optimal Gating for Single-Photon Detectors}{Frederic J. N. Jorgensen and Steven G. Johnson}

\newcommand{\paperfunding}{This work was supported in part by the
Singapore--MIT Alliance for Research and Technology (SMART) Wafer-scale
Integrated Sensing Devices based on Optoelectronic Metasurfaces (WISDOM)
interdisciplinary research group, and by a grant from the Simons Foundation.}

\title{Optimal Gating for Dead-Time–Constrained Single-Photon Detectors\thanks{Submitted to the editors October 11, 2026.
\funding{\paperfunding}}}

\author{Frederic J. N. Jorgensen\thanks{Department of Mathematics,
Massachusetts Institute of Technology, Cambridge, MA 02139, USA
(\mbox{\email{fjorgen@mit.edu}}, \mbox{\email{stevenj@math.mit.edu}}).}
\and Steven G. Johnson\ifarxiv\footnotemark[1]\else\footnotemark[2]\fi}

\usepackage{amsopn}

%% file: figure1_labels.tex
% Generated by figure_sources/regret.py. Citations remain live in LaTeX.
\begin{tikzpicture}[x=1bp,y=1bp]
  \node[anchor=south west,inner sep=0] at (0,0)
    {\includegraphics{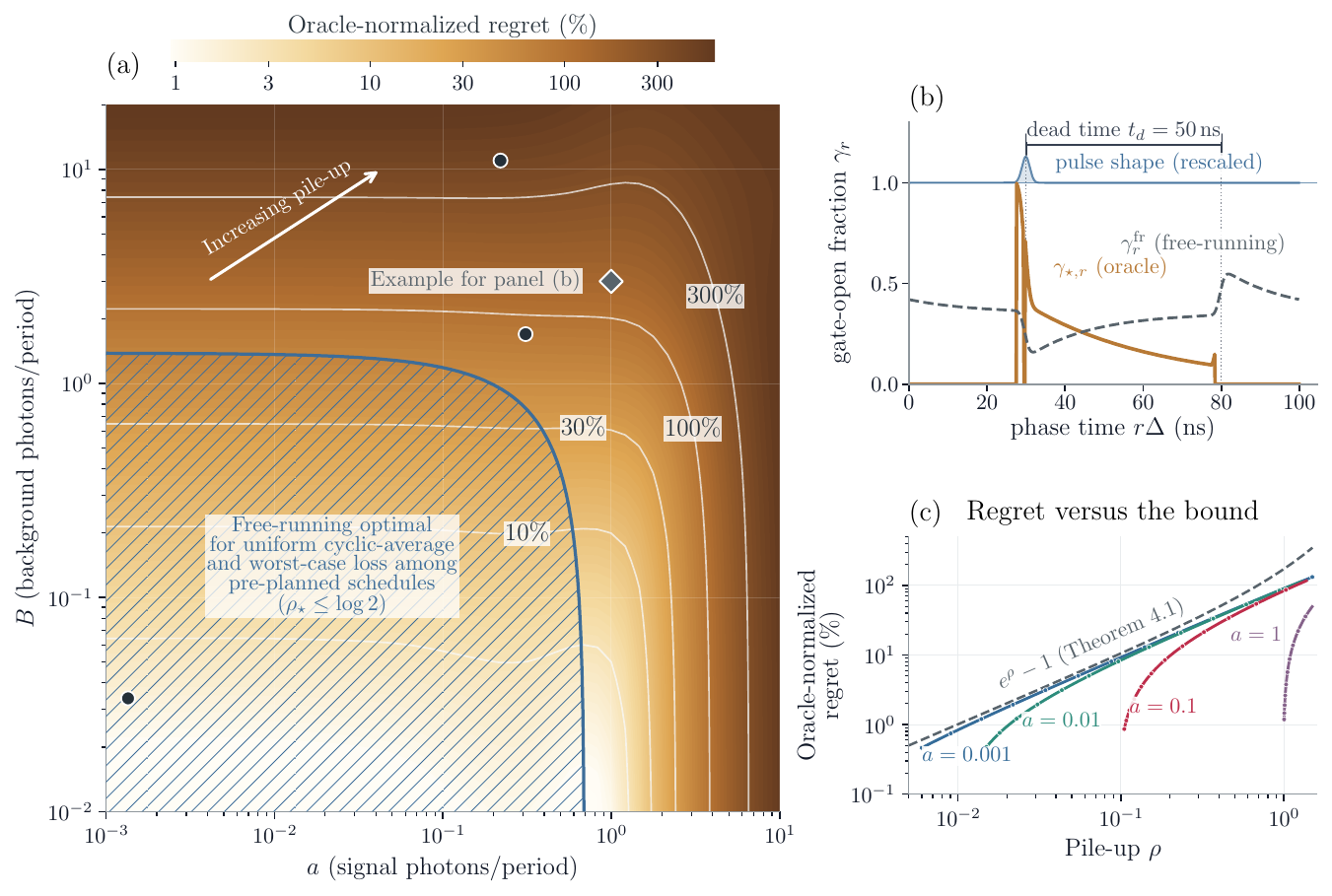}};
  \begin{scope}[every node/.style={font=\fontsize{10}{12}\selectfont,
    text={rgb,255:red,37;green,48;blue,57},fill=white,fill opacity=0.82,
    text opacity=1,inner sep=0.972bp,outer sep=0pt}]
    \node[anchor=west] at (67.939246,103.116521) {\scalebox{1.06049975}{Rapp low-flux~\cite{rapp2017few}}};
    \node[anchor=center] at (255.876828,255.996790) {\scalebox{1.06049975}{Rapp high-flux~\cite{rapp2021high}}};
    \node[anchor=west] at (250.958678,363.740530) {\scalebox{1.06049975}{Gupta high-ambient~\cite{gupta2019asynchronous}}};
  \end{scope}
\end{tikzpicture}%